\documentclass[twocolumn,showpacs,pre]{revtex4}
\usepackage{amsfonts}
\usepackage{amssymb}
\usepackage{amsmath}
\usepackage{graphicx}

\begin{document}
\title{Circular dielectric cavity and its deformations}
\author{ R. Dubertrand$^1$,  E. Bogomolny$^1$, N.
  Djellali$^{2}$, M. Lebental$^{1,2}$, and \fbox{C. Schmit$^1$}}
 \affiliation{$^1$ Universit\'e
Paris Sud, CNRS UMR 8626,\\ Laboratoire de Physique Th\'eorique et
Mod\`eles Statistiques, 91405 Orsay, France\\
$^2$Ecole Normale Sup\'erieure de Cachan, CNRS UMR 8537, Laboratoire
de Photonique Quantique et Mol\'eculaire, 94235 Cachan, France}
\email{remy.dubertrand@lptms.u-psud.fr}
\date{\today}

\begin{abstract}
The construction of perturbation series for slightly deformed
dielectric circular cavity is discussed in details. The obtained
formulae are checked on the example of  cut disks. A good
agreement is found with direct numerical simulations and far-field
experiments.
\end{abstract}
\pacs{42.55.Sa, 05.45.Mt, 03.65.Sq}
\maketitle

\section{Introduction}

\begin{figure}
\includegraphics[width=.5\linewidth]{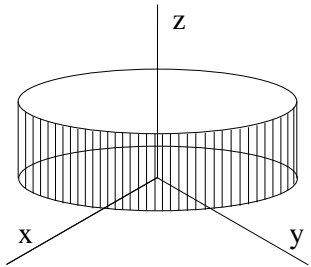}
\caption{Schematic representation of a dielectric cavity.}
\label{cavity}
\end{figure}

Dielectric micro-cavities are now widely used as micro-resonators and
micro-lasers in different  physical, chemical and biological applications
(see e.g. \cite{vahala}, \cite{krioukov} and references therein).
The principal object of these studies is the optical emission from thin
dielectric  micro-cavities of different shapes \cite{stonescience}.
Schematically  such cavity can
be represented as a cylinder whose height is small in comparison with
its transverse dimensions (see Fig.~\ref{cavity}).
If the refractive index of the cavity is $n_1$ and the cavity is surrounded by
a material with the refractive index $n_2<n_1$ (we assume that the
permeabilities in both media are the same) the time-independent Maxwell's equations
take the form (see e.g.  \cite{jackson})
\begin{eqnarray}
\vec{\nabla} \cdot \vec{B}_j=0\ ,&& \vec{\nabla} \cdot n_j^2 \vec{E}_j=0\ ,\nonumber\\
\vec{\nabla} \times \vec{B}_j=-in_j^2 k\vec{E}_j\ ,&&\vec{\nabla} \times
\vec{E}_j=ik\vec{B}_j
\end{eqnarray}
where the subscript $j=1$ (resp.  $j=2$) denotes points inside
(resp. outside) the cavity and $k$ is the wave vector in the vacuum.
These equations have to be completed by the boundary conditions which
follow from the continuity of normal $\vec{B}_{\nu}$ and $n\vec{E}_{\nu}$
and tangential $\vec{E}_{\tau}$ and $\vec{B}_{\tau}$ components
\begin{equation*}
n_1^2\vec{E}_{1\nu}=n_2^2\vec{E}_{2\nu}\ ,\ \vec{B}_{1\nu}=\vec{B}_{2\nu}\ ,\
\vec{E}_{1\tau}=\vec{E}_{2\tau}\ , \ \vec{B}_{1\tau}=\vec{B}_{2\tau}\ .
\end{equation*}
In the true cylindrical geometry, the $z$-dependence of
electromagnetic fields is pure exponential: $\sim {\rm e}^{ {\rm i}
q z}$. Then the above Maxwell  equations can be reduced to the
2-dimensional Helmholtz equations  for the electric field, $E_{jz}$,
and the magnetic field $B_{jz}$ along the axe of the cylinder
\begin{equation}
(\Delta +\tilde{n}_j^2 k^{2})~E_{jz}(x,y)=0\ , \
(\Delta +\tilde{n}_j^2 k^{2})~B_{jz}(x,y)=0
\label{helmholtz}
\end{equation}
with the following boundary conditions
\begin{equation*}
E_{1z}=E_{2z},  B_{1z}=B_{2z},
\frac{\partial E_{1z}}{\partial \tau}=\frac{\partial E_{2z}}{\partial
\tau},
\frac{\partial B_{1z}}{\partial \tau}=\frac{\partial B_{2z}}{\partial \tau},
\end{equation*}
and
\begin{eqnarray}
\frac{1}{\tilde{n}_1^2}\frac{\partial B_{1z}}{\partial \nu}-
\frac{1}{\tilde{n}_2^2}\frac{\partial B_{2z}}{\partial \nu}&=&
\frac{q(n_2^2-n_1^2)}{k\tilde{n}_1^2\tilde{n}_2^2}
\frac{\partial E_{z}}{\partial \tau}\ , \nonumber\\
\frac{n_1^2}{\tilde{n}_1^2}\frac{\partial E_{1z}}{\partial \nu}-
\frac{n_2^2}{\tilde{n}_2^2}\frac{\partial E_{2z}}{\partial \nu}&=&
\frac{q(n_2^2-n_1^2)}{k\tilde{n}_1^2\tilde{n}_2^2}
\frac{\partial B_{z}}{\partial \tau}\ .
\label{bc}
\end{eqnarray}
Here $\tilde{n}_j^2=n_j^2-q^2/k^{2}$
plays the role of the effective two-dimensional (in the $x-y$ plane) refractive index.

When fields are independent on $z$ (i.e. $q=0$) boundary conditions
(\ref{bc}) do not mix $B_z$ and $E_z$ and the two polarizations are
decoupled. They are called transverse electric (TE) field when
$E_z=0$ and transverse magnetic (TM) field  when $B_z=0$. Both cases are
described by the scalar equations
\begin{equation}
(\Delta +\tilde{n}_j^2 k^{2})~\Psi_{j}(x,y)=0
\label{scalar}
\end{equation}
where $\Psi(x,y)$ stands for electric (TM) or magnetic (TE) fields with the
following conditions on the interface between both media:  $\Psi_1=\Psi_2$ and
\begin{eqnarray}
\frac{\partial \Psi_1}{\partial \nu}=\frac{\partial \Psi_2}{\partial \nu} & &\mbox{ for TM polarization}\ ,\label{TM}\\
\frac{1}{n_1^2}\frac{\partial \Psi_1}{\partial \nu}=\frac{1}{n_2^2}
\frac{\partial \Psi_2}{\partial \nu} & & \mbox{ for TE polarization} \ .\label{TE}
\end{eqnarray}
These equations are, strictly speaking, valid only for an infinite
cylinder but they are widely used for a thin dielectric cavities by
introducing the effective refractive index corresponding to the
propagation of confined modes in the bulk of the cavity (see e.g.
\cite{cavity}). In practice, it reduces to small changes in the
refractive indices (which nevertheless is of importance for careful
comparison with experiment \cite{polygons}). For simplicity we will
consider below two-dimensional equations (\ref{scalar}) as the exact
ones.

Only in very limited cases, these equations can be solved
analytically. The most known case is the circular cavity (the disk)
where variables are separated in polar coordinates. For other cavity
shapes tedious numerical simulations are necessary.

The purpose of this paper is to develop perturbation series for
quasi-stationary spectrum and corresponding wave functions for
general cavities which are small deformations of the disk. The
obtained formulae are valid when an expansion parameter is small
enough. The simplicity, the generality, and the physical
transparency of the results make such approach of importance for
technological and experimental applications.

The plan of the paper is the following. In Section~\ref{disc} the
calculation of quasi-stationary states for a circular cavity is
reviewed for completeness. Special attention is given to certain
properties rarely mentioned in the literature. The construction of
perturbation series for eigenvalues and eigenfunctions of small
perturbations of circular cavity boundary is discussed in
Section~\ref{perturbation}. The conditions of applicability of
perturbation expansions are discussed in Section~\ref{degenerate}.
The obtained general formulae are then applied to the case of cut
disks in Section~\ref{cut}. Some technical details are collected in
Appendices.

\section{Dielectric disk}\label{disc}

Let us consider a  two dimensional circular  cavity of radius $R$
made of a material with $n>1$ refractive index. The region outside
the cavity is assumed to be the air with a refractive index of one.
The two-dimensional equations (\ref{scalar}) for this cavity are
\begin{eqnarray}
(\Delta + n^2 k^2 ) \Psi&=&0\; \mbox{ when } r\le R\ ,\nonumber\\
(\Delta + k^2 ) \Psi&=&0\;  \mbox{ when } r>R\ .
\label{eqdiskdiel}
\end{eqnarray}
There is no true bound states for dielectric cavities. The physical origin of the existence of long lived quasi-bound states is the total internal reflection of rays with the incidence angle bigger than the critical angle 
\begin{equation}
 \theta_c=\arcsin\frac{1}{n}\ .
\label{critical}
\end{equation}
To investigate quasi-bound states one imposes outgoing boundary condition at infinity, namely, we require that far
from the cavity there exist only outgoing waves
\begin{equation*}
  \Psi(\vec{x}\ )\propto {\rm e}^{{\rm i} k |\vec{x}\,|}\; \mbox{ when } |\vec{x}\,| \to \infty\ .
\end{equation*}
In cylindrical coordinates $(r,\theta)$, the general form of the
solutions is the following
\begin{equation}
  \label{soldiskdiel}
  \Psi(r,\theta)=\left\{
\begin{array}{ll}
a_m J_m(nkr) {\rm e}^{{\rm i} m \theta}\ , & r \le R\ ,\\
b_m H_m^{(1)}(kr) {\rm e}^{{\rm i} m \theta}\ , & r >R\ ,
\end{array}\right.
\end{equation}
where $m=0,1,\ldots $ is an integer (the azimuthal quantum number)
related to the orbital momentum. $J_m(x)$ (resp.$H_m^{(1)}(x)$)
stands for the Bessel function (resp. the Hankel function of the
first kind) of order $m$. Due to rotational symmetry, eigenvalues
with $m\neq 0$ are doubly degenerated.

By imposing the boundary conditions (\ref{TM}) or (\ref{TE}) one gets the quantization condition
\begin{equation}
  \label{reldisp}
   \frac{n}{\nu}\frac{J_m'}{J_m}(nk R)=\frac{H_m^{(1)\prime}}{H_m^{(1)}}(k R)
\end{equation}
where
\begin{equation}
\nu=\left \{\begin{array}{cc} 1&\mbox{ for TM polarization}\\
n^2&\mbox{ for TE polarization}
\end{array}  \right .\ .
\label{nu}
\end{equation}
The quasi-stationary eigenvalues of this problem depend on azimuthal
quantum number $m$ and on other quantum number $p$ related with
radial momentum: $k=k_{m p}$. They are  complex numbers
\begin{equation}
  k=k_r+{\rm i}k_i
\end{equation}
where $k_r$ determines the position of a resonance and $k_i<0$ is related with its lifetime.

\begin{figure*}
\begin{minipage}{.49\linewidth}
\includegraphics[width=.9\linewidth, angle=-90]{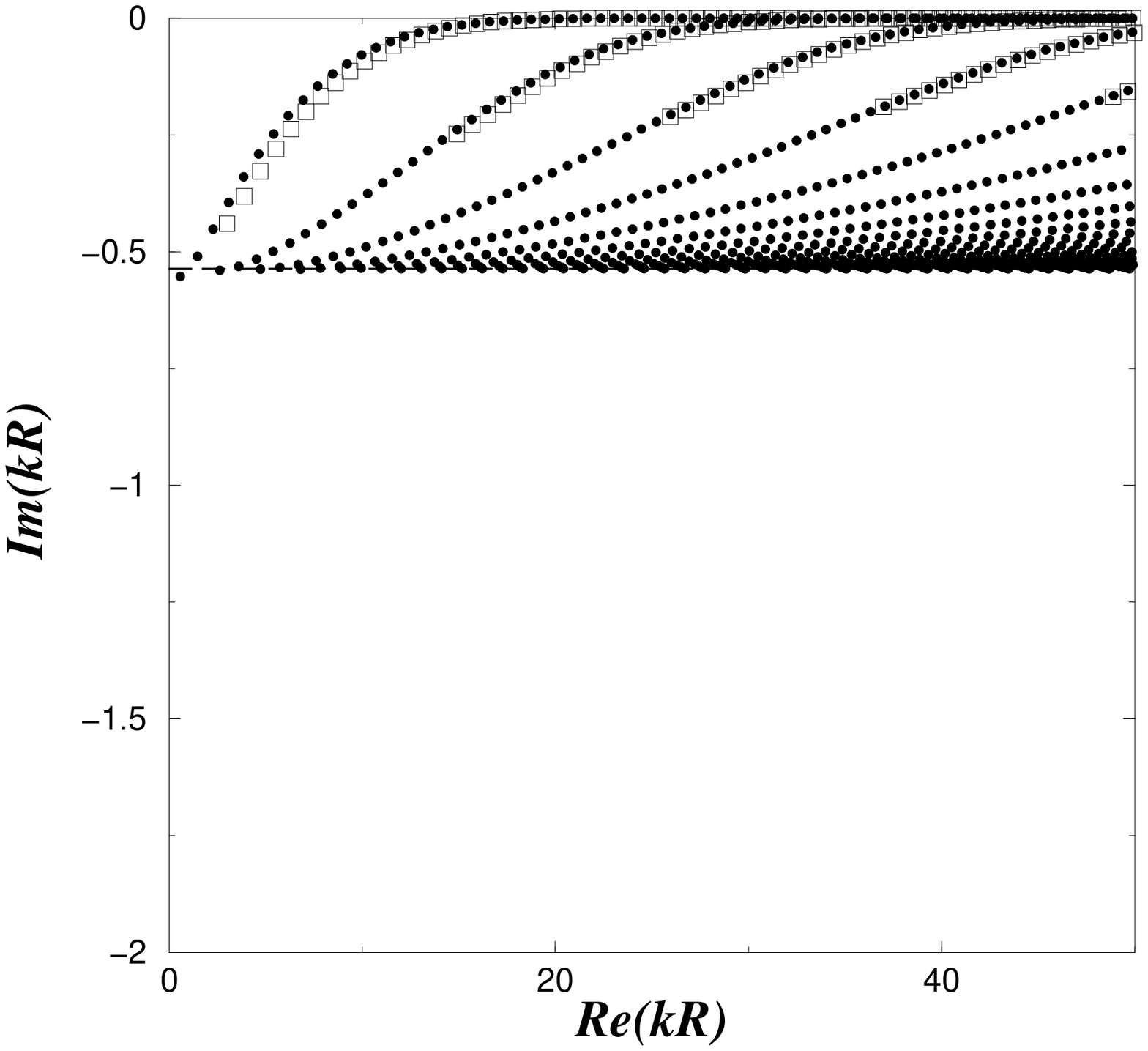}
\begin{center}a)\end{center}
\end{minipage}\hfill
\begin{minipage}{.49\linewidth}
\includegraphics[width=.9\linewidth, angle=-90]{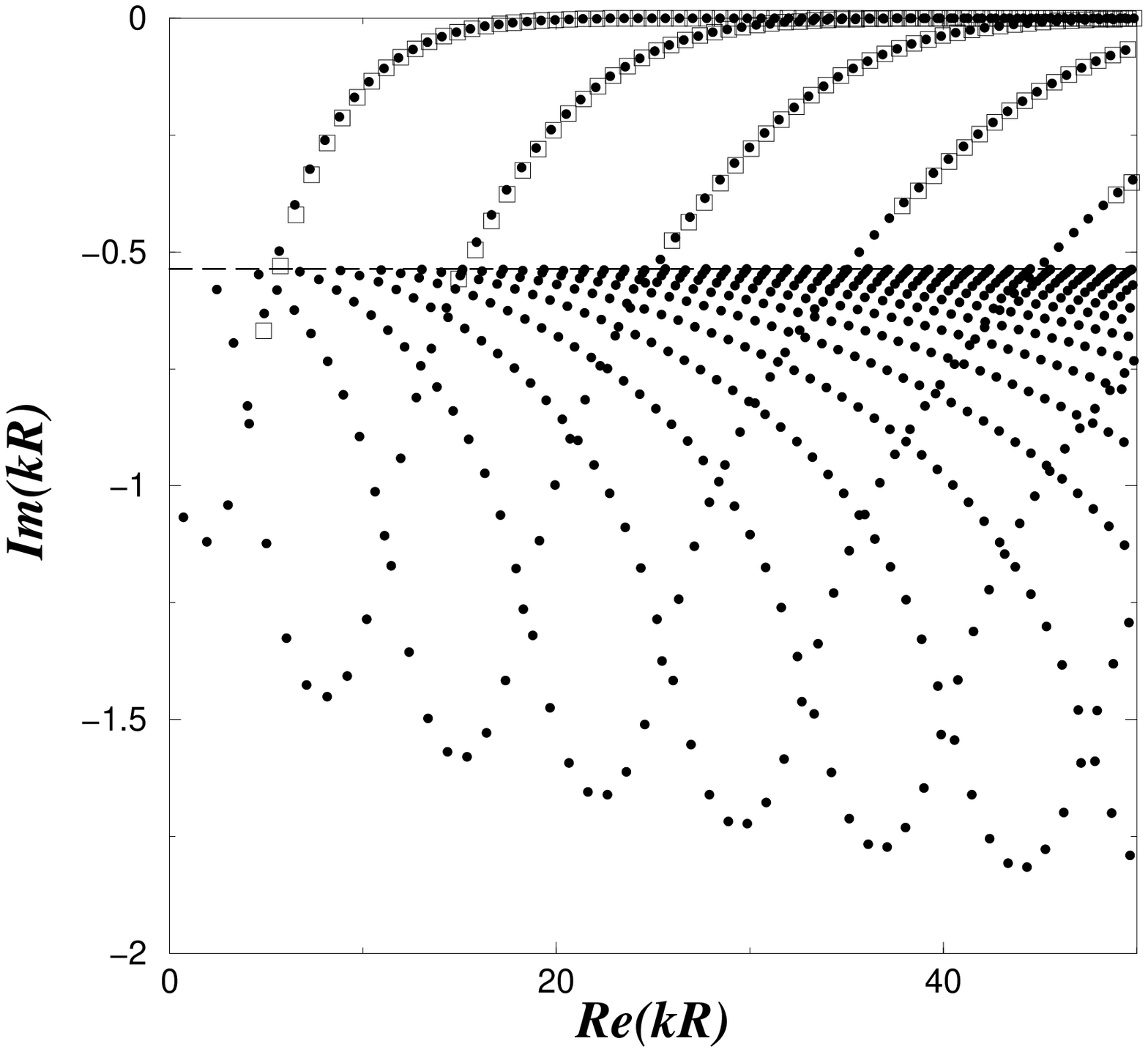}
\begin{center}b)\end{center}
\end{minipage}
\caption{Quasi-stationary eigenvalues for a circular cavity
  with $n=1.5$. a) TM polarization, b)  TE polarization. Filled
  circles are deduced from direct numerical resolution of Eq.
  (\ref{reldisp}), while open squares indicate semiclassical approximation
  for these eigenvalues based
on Eqs.~(\ref{reel_scl}) and (\ref{imag_scl}) when $|kR|<m<n|kR|$. }
\label{nivo_ex}
\end{figure*}

In Fig.~\ref{nivo_ex} we plot solutions of Eq.~(\ref{reldisp})
obtained numerically for a cylindrical cavity with refractive index
$n=1.5$. Points are organized in families corresponding to different
values of radial quantum number $p$.
The dotted line in these figures indicates the classical lifetime of
modes with fixed $m$ and $k\to\infty$. Physically these modes correspond to
waves propagating along the diameter  whose lifetime is given by
\begin{equation}
{\rm Im }(kR)=\frac{1}{2n}\ln\left(\frac{n-1}{n+1}\right)\simeq -0.53648
\label{minimal}
\end{equation}
In the semiclassical limit and for Im$(kR)\ll$ Re$(kR)$ simple
approximate formulae can be obtained from standard approximation of
Bessel and Hankel functions  \cite{bateman}:
\begin{itemize}
 \item when  $m<z$
\begin{eqnarray}
&&J_m(z)=\sqrt{\frac{2}{\pi}}\frac{1}{(z^2-m^2 )^{1/4}} \nonumber\\
&\times& \cos\left( \sqrt{z^2-m^2}- m \arccos{\frac{m}{z}} -\frac{\pi}{4} \right)\ ,
\label{Bessel}
\end{eqnarray}
\item when  $m>z$
\begin{eqnarray}
&&H_m^{(1)}(z)= -{\rm i }  \sqrt{\frac{2}{\pi}}
\frac{{\rm e}^{-\sqrt{m^2-z^2}}}{(m^2-z^2)^{1/4}} \nonumber \\
&\times&\left(\frac{m}{z} +\sqrt{\left(\frac{m}{z}\right)^2-1}\right)^{m}\ .
\label{Hankel}
\end{eqnarray}
\end{itemize}
Denoting $x={\rm Re}(kR)$ and $y={\rm Im}(kR)$ and assuming that $y\ll x$,
$x\gg 1$, and $m/n<x<m$ one gets (see e.g. \cite{nockel}) that the real part
of (\ref{reldisp}) can be transformed to the following form
\begin{eqnarray}
 &&\sqrt{n^2 x^2-m^2}- m \arccos{\frac{m}{nx}} -\frac{\pi}{4} \nonumber\\
&&=  \arctan\nu \sqrt{ \frac{m^2-x^2}{n^2 x^2 - m^2}}+(p-1)\pi
\label{reel_scl}
\end{eqnarray}
where the integer $p=1,2,\ldots$ is the radial quantum number and $\nu$ is defined in
(\ref{nu}).
The imaginary part of equation (\ref{reldisp}) is then reduced to \
\begin{equation}
y\approx -\frac{2}{\pi x (n^2-1) |H_m^{(1)}(x)|^2}\zeta
\label{imag_scl}
\end{equation}
where $\zeta=1$ for TM waves and $\zeta=n^2 x^2/(m^2(n^2+1)-n^2 x^2)$ for TE waves.
When $x$ and $m$ are large, $y$  is
exponentially small as it follows from (\ref{Hankel}).

The above equations can not be applied for the most confined levels
(similar to the ``whispering gallery'' modes in closed billiards)
for which $nx$ is close to $m$.  In Appendix~\ref{whispering}  it is
shown that real part of such quasi-stationary eigenvalues with
${\cal O}(m^{-1})$ precision is given by the following expression
\begin{eqnarray}
&&x_{m,p}=\frac{m}{n}+\frac{\eta_{p}}{n}\Big (\frac{m}{2}\Big )^{1/3}-
\frac{1}{\nu\sqrt{n^2-1}}\label{bottom}\\
&&+\frac{3\eta_{p}^2}{20 n}\Big ( \frac{2}{m}\Big )^{1/3}+
\frac{n^2\eta_{p}}{2\nu (n^2-1)^{3/2}}\Big ( \frac{2}{3\nu^2}-1\Big )
\Big (\frac{2}{m}\Big )^{2/3}\nonumber
\end{eqnarray}
where $\eta_p$ is the modulus of the $p^{\mbox{th}}$ zero of the Airy function (\ref{airy}).

\begin{figure}
\includegraphics[angle=-90, width=.9\linewidth]{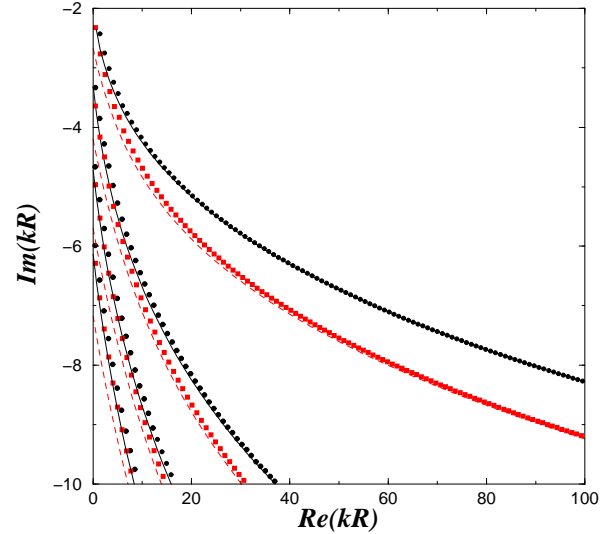}
\caption{Additional branches of quasi-stationary eigenvalues for a
circular cavity with $n=1.5$. Black circles indicate the TM modes
and red squares show
  the position of the TE modes. Solid black and dashed red lines represent the asymptotic result
  (\ref{evas})  for  respectively the TM and the TE modes.}
\label{bizar}
\end{figure}

A more careful study of (\ref{reldisp}) reveals  that there exist other
branches of eigenvalues with large imaginary part not visible in Fig.~\ref{nivo_ex}.
Some of them  are indicated in Fig.~\ref{bizar}.
These states can be called external whispering gallery modes as their wave
functions are practically zero inside the circle. So they are of minor
importance for our purposes. They can also be identified with above-barrier
resonances. In Appendix~\ref{whispering} it is shown that 
in semiclassical limit these states are related with complex zeros of the
Hankel functions and they  are well described asymptotically (with ${\cal
  O}(m^{-1})$ error) as follows 
\begin{eqnarray}
&& x_{m,p}= m+\left(\frac{m}{2}\right)^{1/3} \eta_{p}\ {\rm e}^{-2{\rm i}\pi/3} -
  \frac{{\rm i}\nu}{\sqrt{n^2-1}}\nonumber\\
&&+\frac{3{\rm e}^{-4{\rm i}\pi/3}\eta_{p}^2 }{20}
\left ( \frac{2}{m}\right )^{1/3} \nonumber\\
&&+ \frac{{\rm i}\nu \eta_{p}{\rm e}^{-2{\rm i}\pi/3}}{2
(n^2-1)^{3/2}}\left ( 1-\frac{2}{3}\nu^2\right ) \left
(\frac{2}{m}\right )^{2/3}\label{evas}
\end{eqnarray}
with the same $\eta_p$ as in (\ref{bottom}).

Similar equations have been obtained in \cite{henning}.

\section{Perturbation treatment of deformed circular cavities}\label{perturbation}

\begin{figure}
\begin{center}
\includegraphics[width=.5\linewidth]{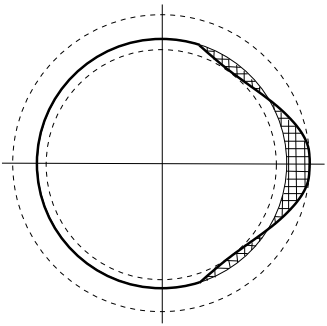}
\caption{Example of a deformed circular cavity. Shaded areas
represent regions where the refractive index differs from the one of
the circular cavity.  } \label{deformation}
\end{center}
\end{figure}

In the previous Section we have considered the  case of a dielectric
circular cavity. It is one of the rare cases of integrable
dielectric cavities in two dimensions. The purpose of this Section
is to develop a perturbation treatment for a general cavity shape
which is a small deformation of the circle (see
Fig.\ref{deformation}).
We consider a cavity which boundary is defined as
\begin{equation}
r=R+\lambda f(\theta)
\label{general}
\end{equation}
in the polar coordinates $(r,\theta)$. Here $\lambda$ is a formal
small parameter aiming at arranging perturbation series.

Our main assumption is that the deformation function $ \lambda f(\theta) $ is small
\begin{equation}
|\lambda  f(\theta)|\ll R\ .
\label{small_classical}
\end{equation}
Of course, for the quantum mechanical perturbation theory this condition is not enough. It is quite natural (and will be demonstrated below) that the criterion of applicability of the quantum perturbation theory is, roughly,
\begin{equation}
 \delta a\  k^2\ll 1
\label{small}
\end{equation}
 where $\delta a$ is the area where perturbation 'potential' $\delta n^2$ is non zero
(represented by dashed regions in Fig.~\ref{deformation}).

To construct the perturbation series for the quasi-stationary
states, we use two complementary methods. In Section
\ref{boundary_conditions} we adapt the method proposed in \cite{yeh,
keller} for diffraction problems. The main idea of this method is to
impose the required boundary conditions (\ref{TM}) or (\ref{TE}) not
along the true boundary of the cavity but on the circle $r=R$. Under
the assumption (\ref{small}) this task can be achieved  by
perturbation series in $\lambda$. In Section \ref{Green_function} we
use a more standard method based on the direct perturbation solution
of the required equations using the Green function of the circular
dielectric cavity. Both methods lead to the same series but they
stress different points and may be useful in different situations.

For clarity we consider only the TM polarization where the field and
its normal derivative are continuous on the dielectric interface.
For the TE polarization the calculations are more tedious but follow
the same steps. To simplify the discussion we assume that the
deformation function $f(\theta)$ is symmetric:
$f(-\theta)=f(\theta)$ (as in Fig.~\ref{deformation}). In this case
the quasi-stationary eigenfunctions are either symmetric or
antisymmetric with respect to this inversion. Then in polar
coordinates, they can be expanded either in $\cos(p\theta)$ or
$\sin(p\theta)$ series. The general case of non-symmetric cavities
is analogous to the case of degenerate  perturbation series and can
be treated correspondingly.

\subsection{Boundary shift}\label{boundary_conditions}

The condition of continuity of the wave function at the dielectric interface states
\begin{equation}
\Psi_1(R+\lambda f(\theta),\theta)=\Psi_2(R+\lambda f(\theta),\theta)
\label{cont_psi}
\end{equation}
where subscripts $1$ and $2$ refer respectively to wave function
inside and outside the cavity. Expanding formally $\Psi_{1,2}$ into
powers of $\lambda$ one gets
\begin{eqnarray}
&&[\Psi_1-\Psi_2](R,\theta)=-
\lambda f(\theta)\left [\frac{\partial \Psi_1}{\partial r}-
\frac{\partial \Psi_2}{\partial r}\right ](R,\theta)\nonumber\\
&-&\frac{1}{2}\lambda^2 f^2(\theta) \left [\frac{\partial^2 \Psi_1}{\partial r^2}-
\frac{\partial^2 \Psi_2}{\partial r^2}\right ](R,\theta)+\ldots \;.
\label{eq1}
\end{eqnarray}
For the TM polarization the conditions (\ref{TM}) imply that the
derivatives of the wave functions inside and outside the cavity
along any direction are the same. Choosing the radial direction, one
gets the second boundary condition
\begin{equation}
\frac{\partial \Psi_1}{\partial r}(R+\lambda f(\theta),\theta)=
\frac{\partial \Psi_2}{\partial r}(R+\lambda f(\theta),\theta)
\end{equation}
which can be expanded over $\lambda$ as follows
\begin{eqnarray}
&&\left [\frac{\partial \Psi_1}{\partial r}-\frac{\partial \Psi_2}{\partial r}\right ](R,\theta)=\nonumber\\
&-&
\lambda f(\theta)\left [\frac{\partial^2 \Psi_1}{\partial r^2}-
\frac{\partial^2 \Psi_2}{\partial r^2}\right ](R,\theta)\label{eq2}\\
&-&\frac{1}{2}\lambda^2 f^2(\theta) \left [\frac{\partial^3 \Psi_1}{\partial r^3}-
\frac{\partial^3 \Psi_2}{\partial r^3}\right ](R,\theta)+\ldots \ .
\nonumber
\end{eqnarray}
We find it convenient to look for the solutions of Eqs.~(\ref{eq1})
and (\ref{eq2}) in the following form
\begin{eqnarray}
\Psi_{1}(r,\theta)&=&\frac{J_m(nkr)}{J_m(nx)}\cos(m\theta)\nonumber\\
&+&\sum_{p\neq m}
a_p\frac{J_p(nkr)}{J_p(nx)}\cos(p\theta)\;,\label{psi_in}\\
\Psi_{2}(r,\theta)&=&(1+b_m)\frac{H_m^{(1)}(kr)}{H_m^{(1)}(x)}\cos(m\theta)\nonumber\\
&+&\sum_{p\neq m}
(a_p+b_p)\frac{H_p^{(1)}(kr)}{H_p^{(1)}(x)}\cos(p\theta)\ .\label{psi_out}
\end{eqnarray}
Here and for all which follows, $x$ stands for $kR$. These
expressions correspond to symmetric eigenfunctions. For
antisymmetric functions all $\cos(\ldots)$ have to be substituted by
$\sin(\ldots)$.

From (\ref{eq1}) and (\ref{eq2}) one concludes  that the unknown coefficients $a_p$,
 and $b_p$ have the following expansions
\begin{equation}
a_p=\lambda \alpha_p+\lambda^2 \beta_p+\ldots\ ,\;\; b_p=\lambda^2
\gamma_p+\ldots\;.
\end{equation}
Correspondingly, the quasi-stationary eigenvalue, $kR\equiv x$, can be represented as
the following series
\begin{equation}
x=x_0+\lambda x_1+\lambda^2 x_2+\ldots\;.
\label{x_series}
\end{equation}
Here $x_0$ is the complex solution of (\ref{reldisp}) which
we rewrite in the form
\begin{equation}
S_m(x_0)=0
\label{qc}
\end{equation}
introducing  for a further use the notation for all $m$ and~$x$
\begin{equation}
S_m(x)=n\frac{J_m'}{J_m}(nx)-\frac{H_m^{(1)\prime}}{H_m^{(1)}}(x)\;.
\label{sm}
\end{equation}
The explicit construction of these perturbation series is presented
in Appendix~\ref{details}. The results are the following. The
perturbed eigenvalue (\ref{x_series}) is
\begin{eqnarray}
x&=&x_0 \Big [ 1-\lambda A_{mm} +\lambda^2 \Big ( \frac{1}{2}(3A_{m m}^2-B_{m
  m})\nonumber\\
&+&x_0(A_{m m}^2 -B_{m m})\frac{H_m^{(1)\prime}}{H_m^{(1)}}(x_0)\label{eigenvalue}\\
&-&(n^2-1)x_0\sum_{k\neq m}A_{m k}\frac{1}{S_k(x_0)}A_{k m}\Big )\Big ] +{\cal O}(\lambda^3)\;.
\nonumber
\end{eqnarray}
The coefficients  of quasi-stationary eigenfunction (\ref{psi_in}) and (\ref{psi_out}) are
\begin{widetext}
\begin{eqnarray}
a_p&=&\lambda x_0(n^2-1)\frac{1}{S_p(x_0)}\Big [A_{p m}+\lambda \Big (
A_{p m}A_{m m}\Big (\frac{x_0}{S_p}\frac{\partial S_p}{\partial x}-1\Big
)\nonumber\\
&+&
\frac{1}{2} B_{p m} \Bigg (1+x_0\Big (\frac{H_m^{(1)\prime}}{H_m^{(1)}}+
\frac{H_p^{(1)\prime}}{H_p^{(1)}}\Big ) \Bigg )
 + x_0(n^2-1)\sum_{k\neq m} A_{p k} \frac{1}{S_k(x_0)} A_{k m} \Big ) \Big ]
 +{\cal O}(\lambda^3)
\label{eigenfunction}
\end{eqnarray}
\end{widetext}
and
\begin{equation}
b_p=\lambda^2\frac{1}{2}x_0^2(n^2-1)B_{p m} +{\cal O}(\lambda^3)
\label{eigenfunction2}
\end{equation}
In these formulae $A_{m n}$ and $B_{m n}$ are the Fourier harmonics
of the perturbation function $f(\theta)$ and its square, given by
(\ref{Amn}) and (\ref{Bmn}) respectively. The above expressions are
quite similar to usual perturbation series and $S_p(x_0)$ plays the
role of the energy denominator.

It is instructive to calculate the imaginary part of the perturbed
level from the knowledge of the first order terms only. Assuming
that Im $(x_0)\ll$ Re $(x_0)$ and using the Wronskian relation (see
e.g. \cite{bateman} 7.11.34)
\begin{equation}
H_m^{(1)\prime}(x)H_m^{(2)}(x)-H_m^{(2)\prime}(x)H_m^{(1)}(x)=\frac{4{\rm
    i}}{\pi x}
\label{wronskian}
\end{equation}
 one gets from (\ref{eigenvalue})
\begin{equation}
{\rm Im}(x)={\rm Im}(x_0) (1+
\rho)-\lambda^2\frac{2x_0(n^2-1)}{\pi}\sum_{p\neq m}\frac{A_{m p}^2}{|S_p H_p^{(1)}|^2}
\label{imaginary_part}
\end{equation}
where (assuming that $m>x_0$ so $H_m^{(1)\prime }/H_m^{(1)}$ is real
cf. (\ref{Hankel}))
\begin{eqnarray}
\rho&=&-\lambda A_{mm}
+\lambda^2 \left [ \frac{1}{2}(3A_{m m}^2-B_{m
  m})\right .\\
&+&\left . x_0(A_{m m}^2 -B_{m m})\frac{H_m^{(1)\prime}}{H_m^{(1)}}(x_0)\right ]\ .
\nonumber
\end{eqnarray}
Expression (\ref{imaginary_part}) without the $\rho$ correction
(which is multiplied by ${\rm Im}(x_0)$ and so negligible for
well-confined levels) can be independently calculated from the
following general considerations. From (\ref{eqdiskdiel}) it follows
that inside the cavity
\begin{equation}
n^2(k^2-k^{*2})\int_{V}|\Psi(x)|^2{\rm d}\vec{x}=J
\end{equation}
where
\begin{equation}
J=\int_B\Big
(\Psi^{*}\frac{\partial}{\partial \vec{\nu}} \Psi
-\Psi \frac{\partial}{\partial \vec{\nu}} \Psi^{*} \Big ){\rm d}\vec{\sigma}\ .
\end{equation}
In the first integral the integration is performed over the volume
of the cavity, $V$, while the second integral is taken over the
boundary of the cavity, $B$. $\nu$ is the coordinate normal to the
cavity boundary and $J$ represents the current through the boundary.
For the TM polarization, it equals the current at infinity.

From (\ref{psi_out}) and (\ref{eigenfunction}) it follows that the field outside the cavity in the first order of the perturbative expansion  is
\begin{eqnarray}
\Psi_{2}(r,\theta)&=&\frac{H_m^{(1)}(kr)}{H_m^{(1)}(x)}\cos(m\theta)\\
&+&\lambda (n^2-1)x_0 \sum_{p\neq m}
A_{p m}\frac{H_p^{(1)}(kr)}{S_p(x)H_p^{(1)}(x)}\cos(p\theta)\ .
\nonumber
\end{eqnarray}
The current  can  be directly calculated from (\ref{wronskian}) or
from the asymptotic of the Hankel functions. Then the current can be
written
\begin{equation}
J=4{\rm i} \Big [\frac{1}{|H_m^{(1)}|^2}+\lambda^2(n^2-1)^2x_0^2
\sum_{p\neq m}
\frac{A_{p m}^2}{|S_pH_m^{(1)}|^2}\Big ] \ .
\end{equation}
To calculate the integral over the cavity volume in the leading
order, the non-perturbed function can be used inside the cavity.
Then the integration over the circle $r=R$ leads to
\begin{equation}
\int_V|\Psi(x)|^2{\rm d}\vec{x}\approx \frac{\pi}{J_m^2(nx_0)} \int_0^R
J_m^2(nkr)r{\rm d}r\ .
\end{equation}
The last integral is (see \cite{bateman} 7.14.1)
\begin{eqnarray*}
&&\int_0^R J_m^2(nkr)r{\rm d}r\\
&&=\frac{R^2}{2}\Big [ J_m^{\prime 2}(nkR)+J_m^2(nkR)(1-\frac{m^2}{(nkR)^2}) \Big ]\ .
\end{eqnarray*}
From the eigenvalue equation (\ref{reldisp}) and the asymptotic
(\ref{Hankel}), it follows that
\begin{equation*}
n\frac{J_m^{\prime}(nkR)}{J_m(nkR)}\approx -\sqrt{m^2/x^2-1} \ .
\end{equation*}
Therefore
\begin{equation}
\int_0^R J_m^2(nkr)r{\rm d}r\approx J_m^2(nx_0)\frac{(n^2-1)R^2}{2n^2}\ .
\end{equation}
Combining these equations leads to
\begin{eqnarray}
{\rm Im }~x&=&-\frac{2}{\pi}\Big [\frac{1}{(n^2-1)x_0 |H_m^{(1)}|^2} \\
&+&\lambda^2(n^2-1)x_0
\sum_{p\neq m}
\frac{A_{p m}^2}{|S_pH_m^{(1)}|^2}\Big ] \;.
\nonumber
\end{eqnarray}
According to (\ref{imag_scl}) the first term of this expression is the imaginary part
of the unperturbed quasi-stationary eigenvalue (assuming that Im $(x_0)\ll$ Re
$(x_0)$)  and one gets (\ref{imaginary_part}) using only the first order
corrections. The missing terms are proportional to the imaginary part of the unperturbed level and  can safely be neglected for the well confined levels.

These calculations clearly demonstrate that the deformation of the
cavity leads to the scattering of the initial well confined wave
function with Re$(kR)<m<n$Re$(kR)$ into all possible  states with
different $p$ momenta. Among these states, some are very little
confined or not confined at all. These states with $p<$Re$(kR)$ give
the dominant contribution to the lifetime of perturbation
eigenstates. Such scattering picture becomes more clear in the Green
function approach discussed in the Section \ref{Green_function}.

The important quantity for applications is the far-field emission.
It is calculated using the coefficients $a_p$ (\ref{eigenfunction})
and $b_p$ (\ref{eigenfunction2}) in (\ref{psi_out}) and substituting
its asymptotic for $H_p^{(1)}(kr)$:
\begin{equation}
H_p^{(1)}(kr)\stackrel{r\to\infty}{\longrightarrow}
\sqrt{\frac{2}{\pi k r}}~{\rm e}^{{\rm i}(kr -\pi p/2-\pi/4)}\ .
\end{equation}
Then one gets
\begin{equation}
 \Psi_2(r,\theta)\stackrel{r\to \infty}{\longrightarrow} \sqrt{\frac{2}{\pi k r}}
~{\rm e}^{{\rm i}(kr -\pi/4)} F(\theta)
\end{equation}
where
\begin{eqnarray}
 F(\theta)&=&(1+b_m)\frac{{\rm e}^{-{\rm i}\pi m/2}}{H_m^{(1)}(x)}\cos(m\theta) \nonumber \\
&+&\sum_{p\neq m}
(a_p+b_p)\frac{{\rm e}^{-{\rm i}\pi p/2}}{H_p^{(1)}(x)}\cos(p\theta)\ .
\label{long_range}
\end{eqnarray}

The boundary shift method discussed in this Section is a simple and
straightforward approach to perturbation series expansions  for
dielectric cavities. As it is based on
Eqs.~(\ref{cont_psi})-(\ref{eq2}), it first shrinks to zero the
regions where the refractive index differs from its value for the
circular cavity. Consequently, the calculation of the field
distribution in these regions remains unclear. Besides the direct
continuation of perturbation series (\ref{eigenfunction}) inside
these regions diverges. To clarify this point, we discuss in the
next Section a different method without such drawback.

\subsection{Green function method}\label{Green_function}

Fields in two-dimensional dielectric cavities obey the Helmholtz
equations (\ref{helmholtz}) which can be written as one equation in
the whole space for TM polarization
\begin{equation}
 \left ( \Delta +k^2n^2(\vec{x}\,)\right )\Psi(\vec{x}\,)=0
\label{schrodinger}
\end{equation}
with position dependent 'potential' $n^2(\vec{x}\,)$. For perturbed cavity (\ref{general})
\begin{equation}
 n^2(\vec{x}\,)=n^2_0(\vec{x}\,)+\delta n^2(\vec{x}\,)
\label{potential}
\end{equation}
where $n^2_0(\vec{x}\,)$ is the 'potential' for the pure circular cavity
\begin{equation}
 n^2_0(\vec{x}\,)=\left \{ \begin{array}{cc} n^2&\mbox{ when }|\vec{x}\,|<R\\1&\mbox{ when }|\vec{x}\,|>R
\end{array}\right . \;.
\label{circular_potential}
\end{equation}
and the perturbation $\delta n^2(\vec{x}\,)$ is equal to
\begin{eqnarray}
&(n^2-1)&\;\mbox{ when }f(\theta)>0 \mbox{ and } R<|\vec{x}\,|< R+\lambda f(\theta)\nonumber \\
&-(n^2-1)&\; \mbox{ when }f(\theta)<0 \mbox{ and } R+\lambda f(\theta)<|\vec{x}\,|< R \nonumber  \\
&0&\;\mbox{ in all other cases }
\label{delta_n}
\end{eqnarray}
Hence, the integral of $\delta n^2(\vec{x}\,)$ with an arbitrary function $F(\vec{x}\,)\equiv F(r,\theta)$ can be calculated as follows
\begin{equation}
 \int \delta n^2(\vec{x}\,)F(\vec{x}\,){\rm d}\vec{x}=(n^2-1)\int {\rm d}\theta
\int_R^{R+\lambda f(\theta)} F(r,\theta)r{\rm d}r\ .
\end{equation}
Eq.~(\ref{schrodinger}) with 'potential' (\ref{potential}) can be rewritten in the form
\begin{equation}
 \left ( \Delta_{\vec{x}} +k^2n^2_0(\vec{x}\,)\right )\Psi(\vec{x}\,)=
-k^2\delta n^2(\vec{x}\,) \Psi(\vec{x}\,)\ .
\end{equation}
Then its formal solution is given by the following integral equation
\begin{equation}
 \Psi(\vec{x}\,)=-k^2\int G(\vec{x},\vec{y}\,)\delta n^2(\vec{y}\,) \Psi(\vec{y}\,){\rm d}\vec{y}   
\label{psi_green}
\end{equation}
where $G(\vec{x},\vec{y}\,)$ is the Green function of the equation for the dielectric circular cavity 
which describes the field produced at point $\vec{x}$ by the delta-function source situated at point $\vec{y}$. The explicit expressions of this function are presented in Appendix~\ref{Green}.

It is convenient to divide the $\vec{x}$ plane into  three circular regions  $r<R_1$, $R_1<r<R_2$, and $r>R_2$ where
\begin{eqnarray}
 R_1&=&\mbox{ min}_{\theta}(R, R+\lambda f(\theta))\nonumber\\
 R_2&=&\mbox{ max}_{\theta}(R, R+\lambda f(\theta))\ .
\label{regions}
\end{eqnarray}
The boundaries of these regions are indicated by dashed circles in
Fig.~\ref{deformation}. Notice that the deformation 'potential'
$\delta n^2(\vec{x}_,)$ is nonzero only in the second region
$R_1<r<R_2$.  Due to singular character of the Green function (cf.
Appendix~\ref{Green}), wave functions inside each region are
represented by different expressions.

Let $(r,\theta)$ be the polar coordinates of point $\vec{x}$. For
simplicity we assume for a moment that $f(\theta)\leq 0$. Using
(\ref{inside}), Eq.~(\ref{psi_green}) in the region $r<R_1$  can be
rewritten in the form
\begin{equation}
 \Psi(\vec{x}\,)= \sum_{p}
\frac{J_p(nkr)}{J_p(nx)}\cos(p\theta)\hat{L}_p[\Psi]
\label{iteration}
\end{equation}
where $\hat{L}_p[\Psi]$ is the following integral operator
\begin{widetext}
\begin{eqnarray}
& & \hat{L}_p[\Psi]=\frac{x^2(n^2-1)}{R^2}\int {\rm d}\phi \cos( p\phi )\label{the_first}\\
&\times&\int_R^{R+\lambda f(\phi)}\rho {\rm d}\rho \left [
\frac{J_p(nk\rho)}{2\pi x S_p(x)J_p(nx)}+
\frac{{\rm i}}{4}(H_p^{(1)}(kn\rho)J_p(nx)-
 H_p^{(1)}(nx)J_p(kn\rho))\right ]\Psi(\rho,\phi) \ .
\nonumber
\end{eqnarray}
\end{widetext}
Assuming that we are looking for corrections to a quasi-stationary state of the non-perturbed circular cavity with the momentum equal $m$, one concludes that 
the quantized  eigen-energies are fixed by the condition that the perturbation terms do not change zeroth order function (see e.g. \cite{morse}), i.e.
\begin{equation}
 \hat{L}_m[\Psi]=1
\label{central}
\end{equation}
which can be transformed into
\begin{eqnarray}
& & S_m(x)=\frac{x^2(n^2-1)}{R^2}\int {\rm d}\phi \cos( p\phi )\nonumber\\
&\times&\int_R^{R+\lambda f(\phi)}\Psi(\rho,\phi) \rho {\rm d}\rho
\left [ \frac{J_m(nk\rho)}{2\pi x J_m(nx)}\right .\label{quantization_condition}\\
&+&\left .\frac{{\rm i} S_m(x)}{4}(H_m^{(1)}(kn\rho)J_m(nx)-
 H_m^{(1)}(nx)J_m(kn\rho))\right ]\ .
\nonumber
\end{eqnarray}
To perform the perturbation iteration of (\ref{iteration}) and
(\ref{quantization_condition}), integrals like the following must be
calculated:
\begin{equation}
 V_{p m}\equiv \frac{1}{J_m(nx)}\hat{L}_p[J_m(kn\rho)\cos(m\phi)]\ .
\end{equation}
For small $\lambda$ the integral over $\rho$ can be computed by expanding the integrand into a series of $\delta r=\rho-R$
\begin{equation}
\int_R^{R+\lambda f(\phi)}F(\rho){\rm d}\rho\approx \lambda f(\phi) F(R)+\frac{1}{2}\lambda^2f^2(\phi)F^{\prime}(R)+\ldots\ .
\label{integral}
\end{equation}
Notice that this method is valid only outside the second region
$R_1<r<R_2$ which shrinks to zero when $\lambda\to 0$ (cf.
(\ref{regions})). In such manner, it leads to
\begin{equation}
 V_{pm}=x^2(n^2-1)(\lambda V_{pm}^{(1)}+\lambda^2 V_{pm}^{(2)})
\end{equation}
where
\begin{equation}
V_{pm}^{(1)}=\frac{1}{xS_p(x)} A_{pm}
\end{equation}
and
\begin{equation}
V_{pm}^{(2)}=\frac{B_{pm}}{2 xS_p(x)}
\big [1+x(\frac{H_p^{(1)\prime}(x)}{H_p^{(1)}(x)}+
\frac{H_m^{(1)\prime}(x)}{H_m^{(1)}(x)})  -2xS_m(x)\big ]\ .
\end{equation}
Here $A_{pm}$ and $B_{pm}$ are defined in (\ref{Amn}) and
(\ref{Bmn}).

The second order terms can also be expressed through $V_{mp}$:
\begin{eqnarray}
\Psi(\vec{x}\,)&=&\frac{J_m(knr)}{J_m(nx)}\cos(m\theta)\\
&+&
\sum_{p\neq m}\frac{J_p(knr)}{J_p(nx)}\cos(p\theta)\Big [V_{pm}+\sum_{k\neq m}V_{p k}V_{km}\Big ]\ .
\nonumber
\end{eqnarray}
The quantization condition (\ref{central}) in the second order states that
\begin{equation}
 V_{mm}+\sum_{k\neq m}V_{m k}V_{km}=1
\end{equation}
which can be expressed as
\begin{eqnarray}
 &&S_m(x)=x(n^2-1)\Big (\lambda A_{mm}+\frac{1}{2}\lambda^2 \big [1 +\\
&&2x\frac{H_m^{(1)\prime}(x)}{H_m^{(1)}(x)}\big ] B_{mm}\Big ) +
\lambda^2 x^2(n^2-1)^2\sum_{k\neq m}
\frac{A_{mk}A_{km}}{S_k(x)}\ .
\nonumber
\end{eqnarray}
Writing as in the previous Section $x=x_0+\lambda x_1+\lambda^2x_2$
where $x_0$ is a zero of $S_m(x)$ and using (\ref{deriv_sp}), one
obtains the same series as  (\ref{eigenvalue}). Other expansions up
to the second order also coincide with the ones presented in the
Section \ref{boundary_conditions}.

To calculate the higher terms of the perturbation expansion, the
wave function must be known in the regions where the perturbation
'potential' $\delta n^2(\vec{x}\,)$ is non-zero. But exactly in
these regions the Green function differs from the one used in
Eq.~(\ref{the_first}). In other words, a method must be found for
the continuation of the wave functions defined in the first region
$r<R_1$ (or in the third one $r>R_2$) into the second region
$R_1<r<R_2$.

The straightforward way of such a continuation is to use explicit formulae for the Green function in the second region and to perform the necessary calculations. As the radial derivative of the Green function is discontinuous, delta-function contributions will appear in certain domains when calculating the integrals as in (\ref{integral}).
One can check that this singular contribution appears in the bulk only in the third order in $\lambda$  in agreement with Eqs.~(\ref{eigenvalue}) and (\ref{eigenfunction}).

The expansion of wave functions into series of the Bessel functions (\ref{iteration}), in general, diverges when $r>R_1$ and the Green function method gives the correct continuation inside this region. Another equivalent method consists in a local expansion of wave function into power series in small deviation from the  boundary of convergence. As the value of the function and its radial derivative are assumed to be known along this boundary ($r=R_1$ in our case) the knowledge of the wave equation inside and outside the cavity determines uniquely the wave function in the both regions.

\section{Applicability of perturbation series}\label{degenerate}

In the previous Section the formal construction of perturbation
series has been performed for quasi-bound states in slightly
deformed dielectric cavities. The purpose of this Section is to
discuss in details the conditions of validity of such an expansion.

From (\ref{Amn}) it follows that the coefficients $A_{p m}$ obey the inequality
\begin{equation}
 |A_{p m}|\leq 2 \xi
\end{equation}
where $\xi$ stands for
\begin{equation}
 \xi=\lambda \int \left |\frac{f(\theta)}{R}\right |{\rm d}\theta\approx \frac{\delta a}{\pi R^2}\ .
\end{equation}
Here $\delta a$ is the surface where the perturbation 'potential' $\delta n^2$ is non-zero and $\pi R^2$ is the full area of the unperturbed circle. The last equality is valid when (\ref{small_classical}) is fulfilled which we always assume.

Consequently, the perturbation formulae can be applied providing
\begin{equation}
\xi (n^2-1)x_0 \langle \left |\frac{1}{S_p(x_0)}\right |\rangle \ll 1 
\label{estimate}
\end{equation}
where $\langle F_p\rangle$ indicates the typical value of $F_p$ and
$x$ stands for $\textrm{Re} (kR)$ at a first approximation.

The usual arguments to estimate this quantity for large $x_0$ are
the following. In the strict semiclassical approximation, states
with a corresponding incident angle larger than the critical angle
have a very small imaginary part and are practically true bound
states. For closed circular cavities, the mean number of states
(counting doublets only once) is given by the Weyl law
\begin{equation}
N(E_j<n^2 k^2)=\frac{A n^2}{8\pi} k^2 +{\cal O}(k)
\end{equation}
where $A=\pi R^2$ is the full billiard area and $n$ is the
refractive index. The latter appears because by definition inside
the cavity   the energy  is $E=n^2 k^2$. For a dielectric circular
cavity with radius $R$, the condition that the incidence  angle is
larger than the critical angle  leads to the following effective
area \cite{braun}
\begin{equation}
A_{eff}(n)=\pi R^2 s_n
\end{equation}
where
\begin{eqnarray}
s_n&=&
\frac{2}{R^2}\int_{R/n}^R(1-\frac{2}{\pi}\arcsin \frac{R}{nr})r{\rm d}r\label{noeckel1}\\
&=&\frac{4}{\pi n^2 x}\int_x^{n x}\sqrt{x^2n^2-m^2}{\rm d}m\label{noeckel2}\\
&=&  1-\frac{2}{\pi}\Big(
\arcsin \frac{1}{n}+ \frac{1}{n}\sqrt{1-\frac{1}{n^2}} \Big )\ .
\nonumber
\end{eqnarray}
Here the first integral (\ref{noeckel1}) corresponds to the
straightforward calculation of phase-space volume such that the
incident angle is larger than the critical one and the second
integral (\ref{noeckel2}) is obtained from (\ref{reel_scl}) taking
into account that $x<m<n x$. For $n=1.5$, $s_n\approx 0.22$.

Consequently, the typical distance between two eigenstates is
\begin{equation}
\delta x \sim \frac{4}{n^2 s_n x}\ .
\end{equation}
The eigen-momenta of non-confining eigenstates have imaginary parts of the order of unity (cf. (\ref{minimal}))
and will be ignored.

As $S_m(x_0)=0$, we estimate that for $p\neq m$
\begin{equation}
\langle \frac{1}{S_p}\rangle \sim \left |\frac{1}{S^{\prime} \delta x}\right |\ .
\end{equation}
Using  (\ref{deriv_sp}) one finds that this value is of order of
\begin{equation}
\langle \frac{1}{S_p}\rangle \sim  C~x
\label{est_s}
\end{equation}
where constant $C\sim 0.25 n^2 s_n/(n^2-1)$. With (\ref{estimate}) it leads to the conclusion that for typical
$S_p$ the criterion of applicability of perturbation series is
\begin{equation}
 s_n\frac{\delta a}{8\pi}k^2n^2\ll 1
\end{equation}
which up to a numerical factor agrees with (\ref{small}).

But this statement is valid only in the mean. If there exist
quasi-degeneracies of the non-perturbed spectrum (i.e. there exist $p$ for
which $1/S_p(x)$ is considerably larger that the estimate (\ref{est_s})) then the standard perturbation treatment
requires modifications. As circular cavities are integrable, the real parts of the strongly confined modes are statistically distributed as the Poisson sequences \cite{berry} and they do have a large number of quasi-degeneracies even for small $k$.
For instance, these are  double quasi-coincidences for the dielectric  circular cavity with $n=1.5$
\begin{eqnarray*}
x_{14,2}=16.7170-0.03895\, {\rm i},&& x_{11,4}=16.6976-0.4695\, {\rm i},\\
x_{15,3}=17.5042-0.37540\, {\rm i},&& x_{12,4}=17.5232-0.4612\, {\rm i},\\
x_{17,4}=21.5715-0.41621\, {\rm i},&& x_{14,5}=21.5106-0.4712\, {\rm i},\\
x_{25,1}=19.4799-0.00254\, {\rm i},&& x_{21,2}=19.4830-0.1211\, {\rm i}.
\end{eqnarray*}
We notice also a triple quasi-coincidence
\begin{eqnarray}
x_{46,1}&=&34.3110-2.2206~10^{-6}\ {\rm i}\, \nonumber\\
x_{41,2}&=&34.3167-0.001982\ {\rm i}\ ,\nonumber \\
x_{37,3}&=&34.317-0.06408\ {\rm i}\ .
\label{triple}
\end{eqnarray}
The existence of these quasi-degeneracies means that the
perturbation series require modifications close to these values of
$kR$ for any small deformations of a circular cavity with $n=1.5$ .

Double quasi-de\-ge\-ne\-ra\-cy is the simplest case because there
is only one eigenvalue of the dielectric circle, with quantum
number, say $p$, which eigenvalue is close to the eigenvalue $x_0$
corresponding to the quantum number $m$.  In such a situation,
instead of the zeroth order equation (\ref{qc}), the system of the
following two equations must be considered
\begin{eqnarray}
S_m(x_0(1+\delta x) ) a_1&=&M_{1 1}a_1+M_{1 2}a_2\ ,\nonumber\\
S_p(x_0(1+\delta x) ) a_2&=&M_{2 1}a_1+M_{2 2}a_2 
\label{two}
\end{eqnarray}
where in the leading order $M_{i j}=x_0(n^2-1)A_{i j}$.

Expanding the $S_p$ functions and using the dominant order
(\ref{deriv_sp}) for $S_p^{\prime}$, the system (\ref{two}) can be
transformed into
\begin{eqnarray}
(s_1-\delta x) a_1&=&A_{1 2}a_2\ ,\nonumber\\
(s_2-\delta x) a_2&=&A_{2 1}a_1 
\label{two_new}
\end{eqnarray}
where
\begin{equation}
s_1=\frac{S_m(x_0)}{x_0(n^2-1)}-A_{1 1}\ ,\;
s_2=\frac{S_p(x_0)}{x_0(n^2-1)}-A_{2 2} \ .
\end{equation}
Our usual choice is $S_m(x_0)=0$ but for symmetry we do not impose
it. The compatibility of the system (\ref{two_new}) leads to the
equation
\begin{equation}
\left |\begin{array}{cc}\delta x-s_1&A_{1 2}\\A_{2 1}&\delta x
    -s_2\end{array}\right |=0\ .
\label{two_equation}
\end{equation}
Its solution which tends to $s_1$ when $A_{1 2}A_{2 1}\to 0$ is
\begin{eqnarray}
&&\delta x=\frac{1}{2}(s_1+s_2)+\frac{1}{2}(s_1-s_2)\sqrt{1+\frac{4A_{1 2}A_{2
      1}}{(s_1-s_2)^2}}\nonumber \\
&&=s_1-\frac{A_{1 2}A_{2 1}}{s_2-s_1}~\frac{2}{1+\sqrt{1+4A_{1 2}A_{2 1}/(s_1-s_2)^2}}\ .
\label{two_perturbation}
\end{eqnarray}
When $A_{1 2}A_{2 1}/(s_1-s_2)^2$ is small, $\delta x$ in
(\ref{two_perturbation}) tends to the usual contribution of the
second order  (\ref{eigenvalue}). Therefore in this approximation,
expression (\ref{eigenvalue}) may be used for all $p$ except the one
which is quasi-degenerate with $x_0$. For this later one, the whole
expression (\ref{two_perturbation}) (without $s_1$) has to be used.
A useful approximation proposed in \cite{denis} consists in taking
the modification (\ref{two_perturbation}) for all the non-degenerate
levels which reduces the numerical calculations.

As the circular billiard is integrable, the probability of having
three and more quasi-degeneracies is not negligible (cf.
(\ref{triple})). The necessary modifications can be performed for
any number of levels but the resulting formulae become cumbersome.
In Appendix~\ref{three_point} we present the formulae for three
quasi-degenerate levels.

\section{Cut disk}\label{cut}

\begin{figure}
\begin{center}
\includegraphics[width=.7\linewidth]{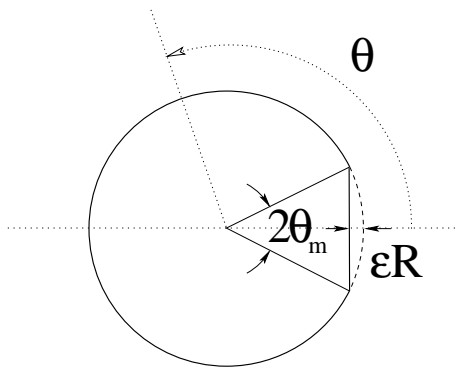}
\end{center}
\caption{Cut disk cavity. $\epsilon R$ is the distance between
  the cut and the circular boundary.}
\label{image_cut_disk}
\end{figure}

As a specific  example, we  consider a deformation of the circular
cavity which is useful for experimental and technological points of
view \cite{doya}. Namely a circle is cut over a straight line (see
Fig.~\ref{image_cut_disk}). Such a deformation is characterized by
the parameter $\epsilon\ll 1$ which determines the distance from the
cut to the circular boundary.
This shape corresponds to the following choice of the deformation
function $f(\theta)$
\begin{equation}
f(\theta) = R\Big (1-\frac{1-\epsilon}{\cos \theta}\Big )\approx R\Big (\epsilon -\frac{\theta^2}{2}+
\frac{\epsilon \theta^2}{2} -\frac{5\theta^4}{24}\Big )
\end{equation}
when  $|\theta|<\theta_m$ and $f(\theta)=0$ for other values of
$\theta$. Here $\theta_m$ is the small angle
\begin{equation}
  \theta_m=\arccos(1-\epsilon)\simeq\sqrt{2\epsilon}+ \frac{\sqrt{2}}{12}
  \epsilon^{3/2}\ .
\label{theta_m}
\end{equation}
Using the formulae discussed in  the preceding Sections, we compute
all the necessary quantities and compare them with the results of
the direct numerical simulations based on a boundary element
representation similar to the one discussed in \cite{wiersig}.

\begin{figure}
\begin{center}
\includegraphics[angle=-90, width=0.9\linewidth]{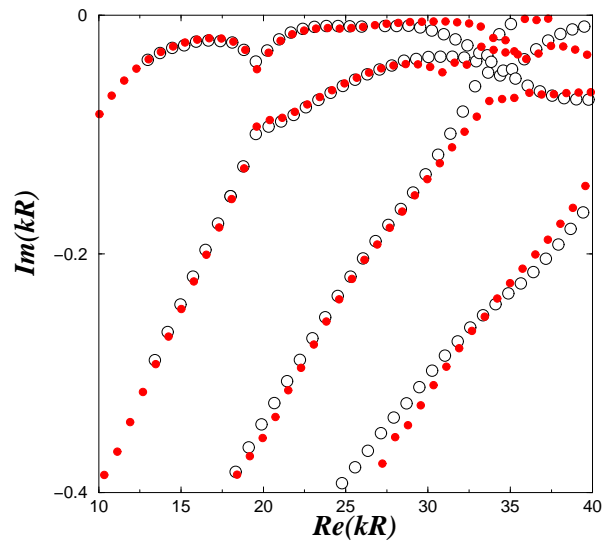}
\end{center}
\caption{Comparison between perturbation and numerical spectra for
the cut disk with $\epsilon=0.05$. Open black circles are results of
the direct numerical simulations and red full circles correspond to
the perturbation expansion (\ref{eigenvalue}).}
\label{spectre-cutdisk}
\end{figure}

The spectrum of quasi-bound states for a cut disk with
$\epsilon=0.05$ is plotted in Fig.~\ref{spectre-cutdisk}. To get a
good agreement in the region $\textrm{Re} (kR)\simeq 15-20$, it is
necessary to take into account double quasi-degeneracies  and, in
the region close to $\textrm{Re} (kR)=35$, triple degeneracy
(\ref{triple}) has been considered. The agreement is quite good even
in the region of large $\textrm kR$ where many families intersect.

\begin{figure*}
\begin{minipage}{.49\linewidth}
\includegraphics[width=.8\linewidth]{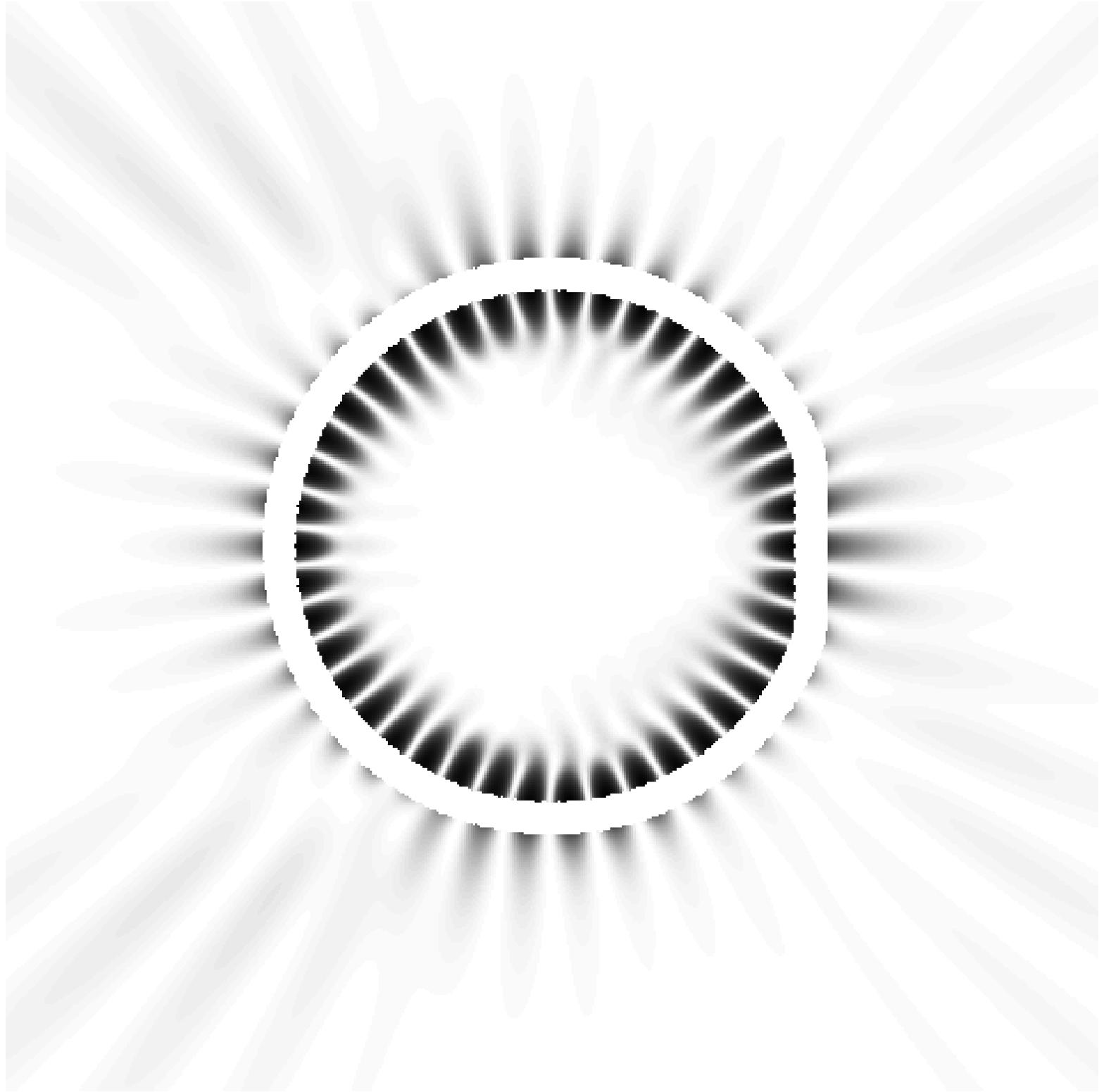}
\begin{center}a)\end{center}
\end{minipage}\hfill
\begin{minipage}{.49\linewidth}
\includegraphics[width=.8\linewidth]{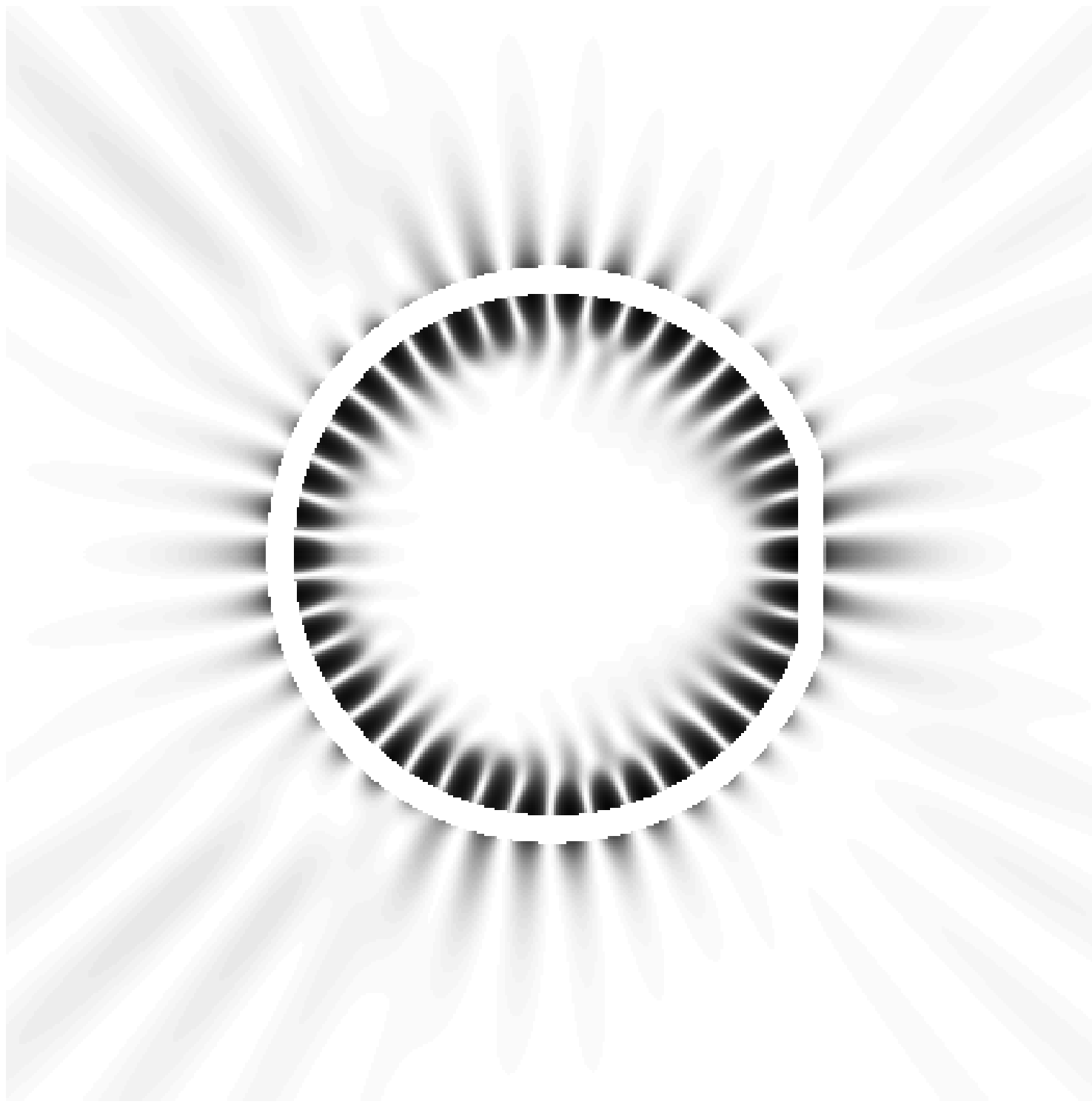}
\begin{center}b)\end{center}
\end{minipage}
\caption{a) Modulus square of the wave function for the cut disk
with $\epsilon=.05$ obtained by direct numerical simulations
corresponding to quasi-stationary eigen-momentum
$kR=16.655-0.0199~{\rm i }$ b) The same but calculated within
perturbation expansion with $m=21$ and $p=0$. The corresponding
eigen-momentum is $kR=16.659-0.0191~{\rm i }$. The high intensity
regions are indicated in black.} \label{comparison}
\end{figure*}

Two wave functions of this cut disk are plotted in
Fig.~\ref{comparison}. The first one is obtained by direct numerical
simulations and the second one corresponds to the perturbation
expansion. Even tiny details are well reproduced by perturbation
computations. 
In the direct numerical simulations, wave functions are
reconstructed from the knowledge of the boundary currents. This
procedure requires the integration of the Hankel function,
$H_0^{(1)}$  \cite{wiersig}. As this function has a logarithmic
singularity, a small region around the cavity boundary has been
removed to reduce numerical errors. It explains the white region in
Fig.~\ref{comparison}a). For an easier comparison, the same region
has been removed also from the perturbation result in
Fig.~\ref{comparison}b).

\begin{figure}
\begin{center}
\includegraphics[width=0.9\linewidth]{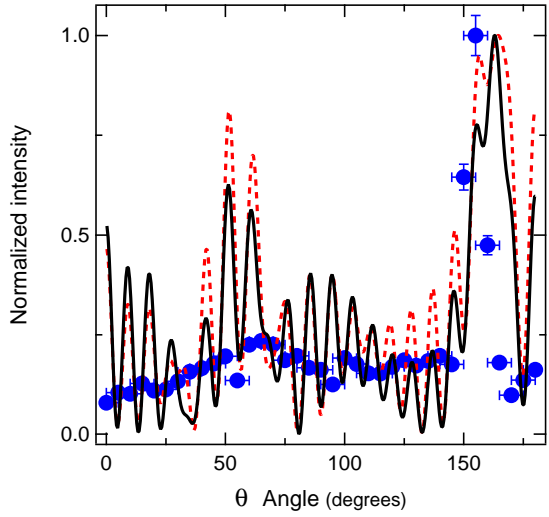}
\end{center}
\caption{Far field emission pattern, $|F(\theta)|^2$, for the same
state as in Fig.~\ref{comparison}. Results from direct numerical
simulations are plotted in solid black line, while the dashed red
line indicates the perturbation result (\ref{long_range}). The blue
points correspond to experimental results with $\epsilon=0.05$ and
$R=60~\mu$m.} \label{farfield-cutdisk}
\end{figure}

\begin{figure}
\begin{center}
\includegraphics[width=0.51\linewidth]{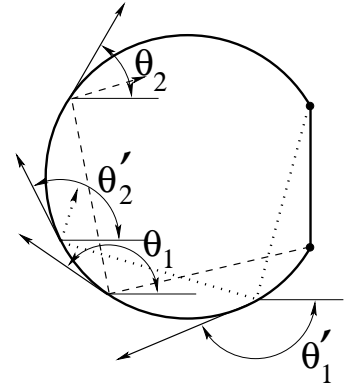}
\end{center}
\caption{Dashed and dotted lines: two main diffracted rays responsible for
  dominant peaks in the far-field  pattern of a dielectric cut disk. All
  rays hit the circular boundary wi th an angle equal the critical angle (\ref{critical}).} 
\label{fig8a}
\end{figure}

Finally, in Fig.~\ref{farfield-cutdisk}, the far-field  emission
pattern  computed numerically for the cut disk is compared with the
same deduced from the perturbation series (\ref{long_range}). The both are normalized to unit maximum. 
Once more a good agreement is found.
The approximate positions of the main peaks in the far-field pattern correspond to the diffracted rays emanated from 
the discontinuities of the cut disk boundaries and reflected at the critical
angle  (\ref{critical}) on the circular boundary (see Fig.~\ref{fig8a}). 
If $(\theta_1, \theta_2,\ldots)$ and $(\theta_1^{\prime},
\theta_2^{\prime},\ldots)$ are the directions of two such refracted rays
one gets from geometrical considerations  
\begin{eqnarray}
 \theta_1&=&\frac{\pi}{2}+2\theta_c-\theta_m\ ,
 \;\theta_1^{\prime}=\frac{3\pi}{2}-2\theta_c-\theta_m\ ,\; \nonumber\\
\theta_2&=&4\theta_c-\theta_m-\frac{\pi}{2}\
,\;\theta_2^{\prime}=4\theta_c+\theta_m-\frac{\pi}{2}\ .
\end{eqnarray}
Here $\theta_c$ is the critical angle (\ref{critical}) and $\theta_m$ is defined in (\ref{theta_m}).  For $\epsilon=.05$ and $n=1.5$, $\theta_1\approx 155.4^{\circ} $, $\theta_1^{\prime}\approx 168.2^{\circ}$, $\theta_2\approx 59^{\circ}$, and $\theta_2^{\prime}\approx 95.4^{\circ}$ which agrees with Fig.~\ref{farfield-cutdisk}. 

\begin{figure}
\begin{center}
\includegraphics[width=0.8\linewidth]{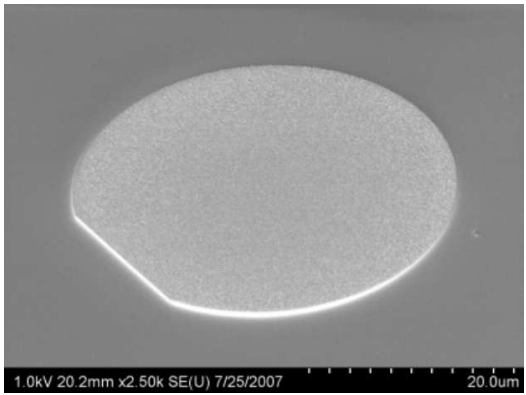}
\end{center}
\caption{Scanning electron microscope image of a cut disk etched
with an electron beam.} \label{photo}
\end{figure}

To complete this study, far-field experiments have been carried out
according to the set-up described in \cite{apl}. The cavities are
made of a layer of polymethylmethacrylate (PMMA) doped by
$4-dicyanomethylene-2-methyl-6-(4-dimethylaminostyryl)-4H-pyran$
(DCM, 5 \% in weight) on a silica on silicon wafer. To obtain a good
resolution of the shape even for such a small cut as
$\epsilon=0.05$, cavities are defined  with  electron beam lithography (Leica
EBPG 5000+) by C. Ulysse (Laboratoire de Photonique et de
Nanostructures, CNRS-UPR20). A scanning electron microscope image of
such a cavity is shown in Fig.~\ref{photo}. 
As specified in
\cite{apl}, the cavities are uniformly pumped one by one from the
top at 532 nm with a pulsed doubled Nd:YAG laser. The light emitted
from the cavity is collected in its plane with a lens leading to a
$10^{\circ}$ apex angle aperture. The directions of emission are
symmetrical about the 0$^{\circ}$ axis according to the obvious
symmetry of the cut-disk shape. In Fig. \ref{farfield-cutdisk}, the
intensity detected in the far-field is plotted versus the $\theta$
angle with blue points. The position of the maximal peak around
$160^{\circ}$ is reproducible from cavity to cavity with
$\epsilon=0.05$ and agrees with both numerical and perturbation
approaches.

The example of the cut disk  clearly demonstrates the usefulness of
the perturbation method presented in this paper for deformed
circular cavities.

\section{Conclusion}

We considered in details the construction of perturbation series for
deformed dielectric circular cavities. The obtained formulae can be
applied for the calculation of the spectrum and the wave functions
as well as other characteristics of these dielectric cavities (e.g.
far-field emission patterns). We checked these formulae on the
example of the cut disk which is of interest from an experimental
point of view. Cavities of other shapes (e.g. spiral \cite{spiral})
can be considered analogously. This method can also be used to
calculate the influence of a small boundary roughness on the
emission properties of circular cavities.

\begin{acknowledgments}
The authors are grateful to J.-S. Lauret and J. Zyss for fruitful
discussions, to D. Bouche for pointing out Ref. \cite{keller}, to
H. Schomerus for pointing out Ref. \cite{henning} and to G.
Faini and C. Ulysse (Laboratoire de Photonique et de Nanostructures,
CNRS-UPR20) for technical support.
\end{acknowledgments}

\appendix

\section{Whispering gallery modes}\label{whispering}

The purpose of this Appendix is to calculate the asymptotic of
quasi-stationary eigenvalues for a fixed radial quantum number $p$
and a large azimuthal quantum number $m\to \infty$.

These resonances are well confined, so let us first find the
corresponding asymptotic expression for zeros of Bessel functions
$J_m(x_m)=0$. To achieve this task it is convenient to use Langer's
formulae (see e.g. \cite{bateman} 7.13.4) which are valid with
${\cal O}(m^{-4/3})$ accuracy
\begin{eqnarray*}
 J_m(x)&=&T(w)[J_{1/3}(z)\cos(\pi/6)-Y_{1/3}(z)\sin(\pi/6)]\ ,
\label{langer}\\
 Y_m(x)&=&T(w) [ J_{1/3}(z)\cos(\pi/6)+Y_{1/3}(z)\sin(\pi/6)]\ ,
\label{langer_Y}
\end{eqnarray*}
where $ T(w)=w^{-1/2}(w-\arctan(w))^{1/2}$, $ w=(x^2/m^2-1)^{1/2}$,
and $z=m(w-\arctan(w))\;.$

The combination of the Bessel functions is expressed as follows
\begin{eqnarray}
&& J_{1/3}(z)\cos(\pi/6)-Y_{1/3}(z)\sin(\pi/6)\nonumber\\
&&=3^{1/6}2^{1/3}z^{-1/3}\textrm{ Ai}\Big (-\big (3 z/2\big
)^{2/3}\Big ) \label{combinaisonlanger}
\end{eqnarray}
where $\textrm{ Ai}(x)$ is the Airy function
\begin{equation}
   \textrm{ Ai}(x)=\frac{1}{\pi}\int_0^{\infty} \cos \big (\frac{t^3}{3}+tx\big ){\rm d} t\ .
\label{airy}
\end{equation}
Therefore in the intermediate region when $z$ is fixed and $m\gg 1$,
the zeros of the $J_m$ Bessel function correspond to
\begin{equation}
 z=\frac{2}{3}\eta_p^{3/2}
\end{equation}
where $\eta_p$ are the modulus of the zeros of the Airy function Ai$(-\eta_p)=0$ ($\eta_1\approx 2.338\;,\;\eta_2\approx 4.088\;,\;\eta_3\approx 5.521$).
Finally, the Bessel function zeros have the following expansion
\begin{equation}
 x_m=m+\alpha m^{1/3}+\beta m^{-1/3}+{\cal O}(m^{-1})
\label{bessel_zeros}
\end{equation}
where
\begin{equation}
 \alpha=2^{-1/3}\eta_p\;,\;\;\beta=\frac{3}{ 2^{5/3}5}\eta_p^2\ .
\label{alpha_beta}
\end{equation}
For $p=1$ this expression agrees numerically with the one given in \cite{bateman} sect. 7.9.

The whispering gallery zeros of Bessel functions
(\ref{bessel_zeros}) permit to calculate explicitly the whispering
gallery modes for the dielectric disk. Indeed we are interesting in
the solutions of equation
\begin{equation}
\label{reldisp2}
 \frac{n}{\nu}  \frac{J_m'}{J_m}(nx)=\frac{H_m^{(1)\prime}}{H_m}(x)
\end{equation}
in the region $xn$ close to $m$, what means
\begin{equation}
 xn=x_m+\delta x
\end{equation}
with $\delta x\ll x_m$. The expansion of the right-hand side of Eq.
(\ref{reldisp2}) leads to
\begin{equation}
 \frac{H_m^{(1)\prime}}{H_m}\big (\frac{x_m}{n}\big )=-\sqrt{n^2-1}+\frac{\alpha n^2}{m^{2/3}\sqrt{n^2-1}}+{\cal O}(m^{-1})\ .
\label{hh}
\end{equation}
In the left-hand side of Eq.~(\ref{reldisp2}),  both numerator and
denominator can be expanded into powers of $\delta x$ taking into
account that $J_m(x_m)=0$
\begin{equation}
 \frac{J_m'}{J_m}(x_m+\delta x)\approx \frac{J_m'+\delta x J_m''+(\delta x)^2J_m'''}
{\delta x J_m'+(\delta x)^2 J_m''/2+(\delta x)^3J_m'''/6}(x_m)\ .
\label{jj}
\end{equation}
Using the Bessel equation
\begin{equation}
 J_m^{\prime \prime}(z)+\frac{1}{z}J_m^{\prime}(z)+\left (1-\frac{m^2}{z^2}\right )J_m(z)=0
\label{bessel_equation}
\end{equation}
one can check that all but one derivatives $J_m^{(k)}/J_m'(x_m)$ are at most ${\cal O}(m^{-1})$ and can be neglected. The only exception is
\begin{equation}
 \frac{J_m'''}{J_m'}(x_m)=-\frac{2\alpha}{m^{2/3}}+{\cal O}(m^{-1})\ .
\end{equation}
Finally expansion (\ref{jj}) takes the form
\begin{equation}
 \frac{J_m'}{J_m}(x_m+\delta x)=\frac{1}{\delta x}-\delta x \frac{2\alpha }{3 m^{2/3}}+{\cal O}(m^{-1})\ .
\label{jprime}
\end{equation}
Combining this equation with  (\ref{reldisp2}) and (\ref{hh}) one
obtains
\begin{equation}
 \delta x=-\frac{n}{\nu \sqrt{n^2-1}}+\frac{\alpha n^3}{m^{2/3}\nu (n^2-1)^{3/2}}(\frac{2}{3\nu^2}-1)+
{\cal O}(m^{-1}).
\end{equation}
These formulae lead to Eq.~(\ref{bottom}).

Other 'whispering gallery' modes correspond to $x$ close to $m$.
With the same notations as above, Langer's formula can be written
for the Hankel function
\begin{equation}
 H_m^{(1)}(x)=T(w){\rm e}^{{\rm i}\pi/6}H_{1/3}^{(1)}(z)+
{\cal O}(m^{-4/3})\ .
\label{langer_H}
\end{equation}
The zeros of $H_{1/3}(z)$ exist only when $z=r{\rm e}^{-{\rm i}\pi}$
with real $r$.   From formula  \cite{bateman} 7.11.42, it follows
that
\begin{eqnarray}
 H_{\nu}^{(1)}(r{\rm e}^{{\rm i}m \pi})&=&-\frac{\sin \pi (m-1)\nu}{\sin \pi \nu}H_{\nu}^{(1)}(r)+\nonumber\\
&+&{\rm e}^{-{\rm i}\pi\nu }\frac{\sin \pi m \nu}{\sin \pi \nu}H_{\nu}^{(2)}(r)\ .
\end{eqnarray}
In this way we get
\begin{eqnarray}
 &&H_{1/3}^{(1)}(r{\rm e}^{-{\rm i}\pi})=2{\rm e}^{-{\rm i}\pi/6}\nonumber\\
&\times& \big [ J_{1/3}(r)\cos(\pi/6)-Y_{1/3}(r)\sin(\pi/6)\big ]\ .
\end{eqnarray}
This is the same combination of the Bessel functions as Eq.
\ref{combinaisonlanger}) for $J_m(x)$. Therefore the first complex
zeros of the Hankel function, $H_m^{(1)}(\tilde{x}_m)$, have a form
similar to (\ref{bessel_zeros})
\begin{equation}
\tilde{x}_m=m+ \tilde{\alpha} m^{1/3}+\tilde{\beta} m^{-1/3}+{\cal O}(m^{-1})
\end{equation}
where $\tilde{\alpha}={\rm e}^{-2\pi{\rm i}/3}\alpha$ and $\tilde{\beta}={\rm e}^{-4\pi{\rm i}/3}\beta$ with the same $\alpha$ and $\beta$ as in (\ref{alpha_beta}).

The next step is to find the asymptotic of $J_m(nx)$ for complex
$x=\tilde{x}_m$. As
\begin{equation}
{\mbox Im}(\tilde{x}_m)=-2^{-1/3}\sin(2\pi/3)\eta_p m^{1/3}+{\cal
O}(m^{-1/3})\ ,
\end{equation}
it tends to $-\infty$ with increasing of $m$. Therefore from
(\ref{Bessel}) it follows that, instead of the $\cos(\ldots)$ term,
only the positive exponent has to be taken into account. Then the
left-hand side of (\ref{reldisp2}) leads to
\begin{eqnarray}
&&\frac{J_m'}{J_m}(n\tilde{x}_m)={\rm i}\sqrt{1-\frac{m^2}{n^2\tilde{x}_m^2}}+{\cal O}(m^{-1}) \label{jjj}\\
&=&{\rm i}\frac{\sqrt{n^2-1}}{n}\left (1+\frac{\tilde{\alpha}}{(n^2-1)m^{2/3}}\right )+{\cal O}(m^{-1}) \ .
\nonumber
\end{eqnarray}
As $H_m^{(1)}(z)$ also obeys  Eq.~(\ref{bessel_equation}) one obtains the same expansion as (\ref{jprime})
\begin{equation}
 \frac{H_m^{(1)\prime}}{H_m^{(1)}}(\tilde{x}_m+\delta x))=\frac{1}{\delta x}-\delta x \frac{2\tilde{\alpha} }{3 m^{2/3}}+{\cal O}(m^{-1})\ .
\label{hprime}
\end{equation}
Eqs.~(\ref{jjj}) and (\ref{hprime}) lead to the following value of $\delta x$ with ${\cal O}(m^{-1})$ accuracy
\begin{equation}
\delta x=-\frac{{\rm i}\nu}{\sqrt{n^2-1}}+\frac{{\rm i}\nu \tilde{\alpha}}{m^{2/3}\sqrt{n^2-1}}
\left (1-\frac{2\nu^2}{3}\right )\ .
\end{equation}
Collecting these equations we get (\ref{evas}).

\section{Construction of perturbation series}\label{details}

From Eqs.~(\ref{eq1}) -- (\ref{psi_out}), the following relations
are valid at the circle $r=R$:
\begin{eqnarray*}
\Psi_1-\Psi_2&=&-\lambda^2\sum_p\gamma_p\cos(p\theta)+{\cal O}(\lambda^3)\,\\
\frac{\partial \Psi_1}{\partial r}-\frac{\partial \Psi_2}{\partial r}&=&
k\left [S_m(x)\cos(m\theta) \right .\nonumber\\
&+&\lambda\sum_{p\neq m}
(\alpha_p+\lambda\beta_p)S_p(x)\cos(p\theta) \\
&+&\left .
\lambda^2\sum_p\gamma_p\frac{H_p^{(1)\prime}}{H_p^{(1)}}(x)\cos(p\theta)
\right]+{\cal O}(\lambda^3)\;,
\nonumber
\end{eqnarray*}
\begin{eqnarray*}
&&\frac{\partial^2 \Psi_1}{\partial r^2}-\frac{\partial^2 \Psi_2}{\partial r^2}=
-\frac{k}{R}\left [(S_m(x)+x(n^2-1))\cos(m\theta)\right .\nonumber\\
&&+\left .\lambda \sum_{p\neq m}
\alpha_p(S_p(x)+x(n^2-1))\cos(p\theta)\right ] +{\cal O}(\lambda^2)\ ,
\end{eqnarray*}
and
\begin{eqnarray*}
&&\frac{\partial^3 \Psi_1}{\partial r^3}-\frac{\partial^3 \Psi_2}{\partial r^3}=\left [
kS_m(x)\left (\frac{m^2+2}{R^2}-k^2n^2\right ) \right.\\
&&-k^3(n^2-1)\frac{H_m^{(1)\prime}}{H_m^{(1)}}(x) \left .+k^2\frac{n^2-1}{R}
\right ]\cos(m\theta)
+{\cal O}(\lambda)\ .\nonumber
\end{eqnarray*}
In the zeroth order $S_m(x_0)=0$. Expanding $S_m(x)$ with $x$ as in
(\ref{x_series}) into a series in $\lambda$, one gets
\begin{eqnarray*}
&&S_m(x_0+\lambda x_1+\lambda^2x_2)=\lambda x_1\frac{\partial S_m}{\partial x}\\
&&+\lambda^2\left (x_2
\frac{\partial S_m}{\partial  x}+\frac{1}{2}x_1^2\frac{\partial^2
  S_m}{\partial x^2}\right )+{\cal O}(\lambda^2) 
\end{eqnarray*}
where all derivatives of $S_m$ are taken at $x=x_0$. These
derivatives are deduced from the Bessel equation
(\ref{bessel_equation})
\begin{eqnarray*}
&&\frac{\partial S_m}{\partial x}(x)=
-(n^2-1)-\frac{1}{x}S_m(x)\\
&&-S_m(x)\left (S_m(x)+2\frac{H_m^{(1)\prime}}{H_m^{(1)}}(x)
\right )\ .
\end{eqnarray*}
In particular, when $x=x_0$
\begin{eqnarray}
\frac{\partial S_m}{\partial x}(x_0)&=&-(n^2-1)\label{deriv_sp}\\
\frac{\partial^2 S_m}{\partial
  x^2}(x_0) &=&(n^2-1)\left
  (\frac{1}{x_0}+2\frac{H_m^{(1)\prime}}{H_m^{(1)}}(x_0)\right )\ .
\nonumber
\end{eqnarray}
In the $\lambda$ first order, it leads to
\begin{eqnarray}
&&-x_1(n^2-1)\cos(m\theta)+\sum_{p\neq m}
\alpha_p S_p(x_0)\cos(p\theta)\nonumber \\
&&=\frac{f(\theta)}{R}x_0(n^2-1) \cos(m\theta)\  .
\label{first_order}
\end{eqnarray}
Coefficients $\alpha_p$ and the first eigenvalue correction, $x_1$,
are determined by comparison of the Fourier harmonics  in both sides of
Eq.~(\ref{first_order})
\begin{equation}
\label{alphp}
\alpha_p=(n^2-1)\frac{x_0}{S_p(x_0)}A_{p m}
\end{equation}
and
\begin{equation}
  \label{e1ord}
x_1=-x_0 A_{m m}
\end{equation}
where $A_{p m}$ are the Fourier harmonics of the deformation function
\begin{equation}
A_{p m}=\frac{\epsilon_p}{\pi R}
\int_0^{\pi}f(\theta)\cos(p\theta)\cos(m\theta){\rm d}\theta\ .
\label{Amn}
\end{equation}
Here
\begin{equation}
\epsilon_p=\left \{ \begin{array}{rl}2 &\mbox{ for }p\neq 0\\ 1&\mbox{ for }p= 0
  \end{array}\right . \ .
\label{epsilon_p}
\end{equation}
In the $\lambda$ second order, it leads to the following two
equations
\begin{equation}
\sum_p\gamma_p\cos(p\theta)=\frac{1}{2}x_0^2(n^2-1)\frac{f^2(\theta)}{R^2}\cos(m\theta)
\label{eq21}
\end{equation}
and
\begin{widetext}
\begin{eqnarray}
&&\left [(n^2-1)\Big (-x_2+\frac{1}{2x_0}x_1^2\Bigg
    (1+2x_0\frac{H_m^{(1)\prime}}{H_m^{(1)}}\Big ) \Big )
+\gamma_m\frac{H_m^{(1)\prime}}{H_m^{(1)}}\right ] \cos(m\theta)
\nonumber\\
&&+\sum_{p\neq m} \left [ x_1\alpha_p\frac{\partial S_p}{\partial x}
+ \beta_p S_p +\gamma_p\frac{H_p^{(1)\prime}}{H_p^{(1)}}\right ]\cos(p\theta) \label{eq22}\\
&&=\frac{f(\theta)}{R}\sum_{p\neq m}
\alpha_p\Big (S_p+x_0(n^2-1)\Big )\cos(p\theta)-\frac{f^2(\theta)}{2R^2}x_0(n^2-1)
\left (1-x_0\frac{H_m^{(1)\prime}}{H_m^{(1)}}\right )\cos(m\theta)\ .\nonumber
\end{eqnarray}
\end{widetext}
Using (\ref{first_order}) the right hand side of equation
(\ref{eq22}) can be rewritten as
\begin{eqnarray*}
&&(n^2-1)\frac{f(\theta)}{R} \left (x_0\sum_{p\neq m}\alpha_p\cos(p\theta) +
x_1\cos(m\theta)\right )\\
&&+(n^2-1)\frac{f^2(\theta)}{2R^2}x_0
\left (1+x_0\frac{H_m^{(1)\prime}}{H_m^{(1)}}\right )\cos(m\theta)\ .
\end{eqnarray*}
Unknown coefficients can be determined by equating the Fourier
harmonics in both parts of equations (\ref{eq21}) and (\ref{eq22}).
From (\ref{eq21}) it follows that for all $p$
\begin{equation}
\gamma_p=\frac{1}{2}x_0^2(n^2-1)B_{p m}\ .
\label{gamp}
\end{equation}
where $B_{p m}$ represents the Fourier harmonics of the square of
the deformation function
\begin{equation}
B_{p m}=\frac{\epsilon_p }{\pi R^2}
\int_0^{\pi}f^2(\theta)\cos(p\theta)\cos(m\theta){\rm d}\theta\ .
\label{Bmn}
\end{equation}
For $p\neq m$ Eq.~(\ref{eq22}) gives
\begin{eqnarray*}
&&\label{eq23} \beta_p S_p+x_1\alpha_p\frac{\partial S_p}{\partial x}+
 \gamma_p\frac{H_p^{(1)\prime}}{H_p^{(1)}}
=(n^2-1)\Big [ x_1 A_{p m}\\
&&+x_0\sum_{k\neq m}\alpha_k  A_{p k}+\frac{1}{2}x_0\Big
  (1+x_0\frac{H_m^{(1)\prime}}{H_m^{(1)}}\Big )B_{p m}\Big ]\ .
\nonumber
\end{eqnarray*}
The $m^{\mbox{th}}$ harmonic of the same equation determines the
second correction to the quasi-stationary eigenvalue
\begin{eqnarray*}
&&x_2=\frac{1}{2x_0}x_1^2\Big
    (1+2x_0\frac{H_m^{(1)\prime}}{H_m^{(1)}}\Big )
 +\frac{\gamma_m H_m^{(1)\prime}}{(n^2-1)H_m^{(1)}}-\\
&&\frac{1}{2}x_0\Big (1+ x_0\frac{H_m^{(1)\prime}}{H_m^{(1)}}\Big )B_{m m}-
x_0\sum_{k\neq m}\alpha_k  A_{m k}-x_1A_{m m}\ .
\label{eq24}
\end{eqnarray*}
Rearranging these equations and using the first order results and
Eq.~(\ref{gamp}), one gets
\begin{eqnarray}
 && \label{e2nd}
  x_2=x_0\Bigg [ \frac{1}{2}(3A_{m m}^2-B_{m
    m})\\
&&+x_0\frac{H_m^{(1)\prime}}{H_m^{(1)}}(A_{m m}^2-B_{m m})-
    \sum_{k\neq m}\alpha_k  A_{m,k}\Bigg ]\ ,
\nonumber
\end{eqnarray}
and
\begin{eqnarray}
 && \label{betap}
 \beta_p =x_0\frac{n^2-1}{S_p}\Bigg [
A_{p m}A_{m m}\Big (\frac{x_0}{S_p}\frac{\partial S_p}{\partial x}-1\Big )\\
&&+\frac{1}{2}B_{p m} \Bigg (1+x_0\Big (\frac{H_m^{(1)\prime}}{H_m^{(1)}}+
\frac{H_p^{(1)\prime}}{H_p^{(1)}}\Big ) \Bigg )
 +\sum_{k\neq m}\alpha_k  A_{p k} \Bigg ]\ .
\nonumber
\end{eqnarray}

\section{Green function for the dielectric circular cavity}\label{Green}

The Green function of the dielectric Helmholtz equation for the circular cavity, $G(\vec{x},\vec{y}\,)$,
is defined as the solution of the following equation
\begin{equation}
\left (\Delta_{\vec{x}} +n^2_0(\vec{x}\, )k^2\right )G(\vec{x},\vec{y}\,) =\delta(\vec{x}-\vec{y}\, ) 
\end{equation}
where $n_0^2$ is the 'potential for the circular cavity defined in (\ref{circular_potential}).

Let us first consider the case when the $\vec{y}$ source point is
inside the circle. In this case, when the $\vec{x}$ point is inside
the cavity, the advanced Green function has the form
\begin{eqnarray}
G(\vec{x},\vec{y}\,)&&=\sum_{m=-\infty}^{\infty}A_mJ_m(nkr){\rm e}^{{\rm i}m(\theta-\phi)}\nonumber \\
&&+\frac{1}{4{\rm i}}H_0^{(1)}(kn|\vec{x}-\vec{y}\,|)
\end{eqnarray}
and when the $\vec{x}$ point is outside the circle
\begin{equation}
G(\vec{x},\vec{y}\,)=
\sum_{m=-\infty}^{\infty}B_m H_m^{(1)}(kr){\rm e}^{{\rm i}m(\theta-\phi)}
\end{equation}
Here and below we assume that points $\vec{x}$ and $\vec{y}$ have
polar coordinates $(r,\theta)$ and $(\rho,\phi)$ respectively.

Constants $A_m$ and $B_m$ are calculated from the boundary
conditions on the interface using the expansion \cite{bateman}
7.15.29 of $H_0^{(1)}(k|\vec{x}-\vec{y}\,|)$
\begin{equation}
 H_0^{(1)}(k|\vec{x}-\vec{y}\,|)=\sum_{m=-\infty}^{\infty}J_m(kr)H_m^{(1)}(k\rho){\rm e}^{{\rm i}m(\theta-\phi)}
\end{equation}
when $r<\rho$ and
\begin{equation}
H_0^{(1)}(k|\vec{x}-\vec{y}\,|)=
\sum_{m=-\infty}^{\infty}H_m^{(1)}(k r)J_m(k\rho ){\rm e}^{{\rm i}m(\theta-\phi)}
\end{equation}
when $r>\rho$.

For the TM polarization the Green function and its normal derivative
at circle  boundary are continuous. That leads to the following
system of equations (with $x=kR$)
\begin{eqnarray*}
A_m J_m(nx)+\frac{1}{4{\rm i}}J_m(nk\rho)H_m^{(1)}(nx)&=&B_mH_m^{(1)}(x)\\
nA_m J_m^{\prime}(nx)+\frac{n}{4{\rm i}}J_m(nk\rho)H_m^{(1)\prime}(nx)&=&B_mH_m^{(1)\prime }(x)\ .
\end{eqnarray*}
Its solutions are
\begin{eqnarray*}
A_m&=&-\frac{H_m^{(1)}(x)}{2\pi x \Delta_m J_m(nx) }J_m(kn\rho)-
\frac{H_m^{(1)}(nx)}{4{\rm i}J_m(nx)}J_m(kn\rho), \\
B_m&=&-\frac{1}{2\pi x \Delta_m}J_m(kn\rho)\ .
\end{eqnarray*}
where
\begin{eqnarray}
 \Delta_m&=&nJ_m^{\prime}(nx) H_m^{(1)}(x)-J_m(nx) H_m^{(1)\prime}(x)\\
&&\equiv  J_m(nx) H_m^{(1)}(x)S_m(x)\ . \nonumber
\end{eqnarray}
In deriving these expressions, the Wronskian (\cite{bateman}
7.11.29) has been used
\begin{equation}
 J_{\nu}(x)H_{\nu}^{(1)\prime}(x)-J_{\nu}^{\prime}(x)H_{\nu}^{(1)}(x)=\frac{2{\rm i}}{\pi x}\ .
\end{equation}
For the $\vec{y}$ source point outside the circle, when $\vec{x}$ is
inside the cavity, then
\begin{equation}
 G(\vec{x},\vec{y}\,)=\sum_{m=-\infty}^{\infty}C_mJ_m(nkr){\rm e}^{{\rm i}m(\theta-\phi)}
\end{equation}
and when $\vec{x}$ is outside the circle, then
\begin{eqnarray}
G(\vec{x},\vec{y}\,)&=& \sum_{m=-\infty}^{\infty}D_m H_m^{(1)}(kr){\rm e}^{{\rm i}m(\theta-\phi)}\\
&+&\frac{1}{4{\rm i}}H_0^{(1)}(k|\vec{x}-\vec{y}\,|)\ .
\nonumber
\end{eqnarray}
Constants $C_m$ and $D_m$ are computed exactly as $A_m$ and $B_m$:
\begin{eqnarray*}
C_m &=&-\frac{1}{2\pi x \Delta_m}H_m^{(1)}(k\rho)\ ,\\
D_m &=&-\frac{J_m(nx)}{2\pi x \Delta_m H_m^{(1)}(x)}H_m^{(1)}(k\rho)-\frac{J_m(x)}{4{\rm i}H_m(x)}H_m^{(1)}(k\rho) \ .
\end{eqnarray*}
The final expressions of the Green function follow from the above
formulae.

 When the $\vec{y}$ source point is inside the circle, the plane is divided into three parts:
1)  $r<\rho$, 2) $\rho<r<R$, 3) $R<r$. Denoting the Green function
in these regions by the corresponding numbers, it can be written as
\begin{equation}
 G_j(\vec{x},\vec{y}\,)=\sum_{p=0}^{\infty}g_p^{(j)}(r,\rho)\cos[p(\theta-\phi)]
\label{inside}
\end{equation}
where
\begin{eqnarray*}
&&g_p^{(1)}(r,\rho) =-\frac{1}{2\pi x}
\epsilon_p\frac{J_p(nkr)J_p(nk\rho)}{J_p^2(nx) S_p(x)}\\
&&+\frac{1}{4{\rm i}}\epsilon_p \frac{J_p(knr)}{J_p(nx)}
[ H_p^{(1)}(kn\rho)J_p(nx)-H_p^{(1)}(nx)J_p(kn\rho)],\\
&&g_p^{(2)}(r,\rho) =-\frac{1}{2\pi x}
\epsilon_p\frac{J_p(nkr)J_p(nk\rho)}{J_p^2(nx) S_p(x)}\\
&&+\frac{1}{4{\rm i}}\epsilon_p \frac{J_p(kn\rho)}{J_p(nx)}
[ H_p^{(1)}(kn r)J_p(nx)-H_p^{(1)}(nx)J_p(kn r)],\\
&&g_p^{(3)}(r,\rho) =-\frac{1}{2\pi x}
\epsilon_p\frac{J_p(nk\rho)H_p^{(1)}(k r )}{J_p(nx)H_p^{(1)}(x)S_p(x)}
\end{eqnarray*}
where $\epsilon_p$ was defined in (\ref{epsilon_p}).

 When point $\vec{y}$ is outside the circle, the plane is divided into three different  regions
1)  $r<R$, 2) $R<r<\rho$, 3) $\rho<r$. With the same notation as
above, the Green function can be written
\begin{equation}
 \tilde{G}_j(\vec{x},\vec{y}\,)=\sum_{p=0}^{\infty}\tilde{g}_p^{(j)}(r,\rho)\cos[p(\theta-\phi)]
\label{outside}
\end{equation}
where
\begin{eqnarray*}
&&\tilde{g}_p^{(1)}(r,\rho) =-\frac{1}{2\pi x}
\epsilon_p\frac{J_p(nkr)H_p^{(1)}(k \rho )}{J_p(nx)H_p^{(1)}(x)S_p(x)}\ ,
\label{tG1}\\
&&\tilde{g}_p^{(2)}(r,\rho) =-\frac{1}{2\pi x}
\epsilon_p\frac{H_p^{(1)}(kr)H_p^{(1)}(k\rho)}{H_p^{(1)2}(x) S_p(x)}
\label{tG2}\\
&&+\frac{1}{4{\rm i}}\epsilon_p \frac{H_p^{(1)}(k\rho)}{H_p^{(1)}(x)}
[ H_p^{(1)}(x)J_p(k r)-H_p^{(1)}(kr)J_p(x)]\ ,
\nonumber\\
&&\tilde{g}_p^{(3)}(r,\rho) =-\frac{1}{2\pi x}
\epsilon_p\frac{H_p^{(1)}(kr)H_p^{(1)}(k\rho)}{H_p^{(1)2}(x) S_p(x)}
\label{tG3}\\
&&+\frac{1}{4{\rm i}}\epsilon_p \frac{H_p^{(1)}(k r)}{H_p^{(1)}(x)}
[ H_p^{(1)}(x)J_p(k \rho)-H_p^{(1)}(k\rho)J_p(x)]\ .
\nonumber
\end{eqnarray*}
Notice that $G_1(\vec{x},\vec{y}\,)=G_2(\vec{y},\vec{x}\,)$, $G_3(\vec{x},\vec{y}\,)=\tilde{G}_1(\vec{y},\vec{x}\,)$,
 and $\tilde{G}_2(\vec{x},\vec{y}\,)=\tilde{G}_3(\vec{y},\vec{x}\,)$. It means that in all cases
 the Green function is symmetric: $G(\vec{x},\vec{y}\,)=G(\vec{y},\vec{x}\,)$ as it should be
 (see e.g. \cite{morse}).

\section{Three degenerate levels}\label{three_point}

For three quasi-degenerate levels, instead of
Eq.~(\ref{two_equation}) one gets the $3\times 3$ determinant
\begin{equation}
\left |\begin{array}{ccc} \delta x-s_1&A_{1 2}&A_{1 3}\\A_{2 1}&\delta x
    -s_2&A_{2 3}\\A_{3 1}&A_{3 2}&\delta x -s_3\end{array}\right |=0
\end{equation}
which leads to the cubic equation
\begin{equation}
(\delta x)^3-\sigma_1(\delta x)^2+(\sigma_2-\alpha)\delta x
-\sigma_3+\beta=0\ .
\label{cubic}
\end{equation}
Here $\sigma_i$ are the elementary symmetric functions of $s_i$
\begin{equation*}
\sigma_1=s_1+s_2+s_3\ ,\;
\sigma_2=s_1s_2+s_2s_3+s_3s_1\ ,\;
\sigma_3=s_1s_2s_3\ .
\end{equation*}
and (because $A_{i j}=A_{j i}$)
\begin{eqnarray*}
\alpha&=&A_{1 2}^2+A_{2 3}^2+A_{3 1}^2\ , \\
\beta&=&2A_{1 2}A_{2 3}A_{3 1}+s_1A_{2 3}^2+s_2A_{3 1}^2+s_3A_{1 2}^2\ .
\end{eqnarray*}
To solve Eq. (\ref{cubic}), the next steps are standard. The
substitution
\begin{equation}
\delta x=y+\frac{1}{3}\sigma_1
\label{y}
\end{equation}
transforms Eq.~(\ref{cubic}) to the reduced form
\begin{equation}
y^3+py+q=0
\label{standard}
\end{equation}
where
\begin{equation}
p=-\frac{1}{3}\sigma_1^2+\sigma_2-\alpha\ ,\;
q=-\frac{2}{27}\sigma_1^3+\frac{1}{3}\sigma_1\sigma_2-\sigma_3+\delta\ .
\label{q}
\end{equation}
with $\delta=\beta -\frac{1}{3}\sigma_1\alpha$.
Finally after the transformation
\begin{equation}
y=z-\frac{p}{3z}
\label{z}
\end{equation}
one gets the equation  $z^3-\frac{p^3}{27z^3}+q=0$
which is a quadratic equation in variable
\begin{equation}
w=z^3\;.
\label{w}
\end{equation}
Its solution is
\begin{equation}
w=-\frac{q}{2}\pm \sqrt{\frac{q^2}{4} +\frac{p^3}{27}}\ .
\label{solution}
\end{equation}
Eqs.~(\ref{y}), (\ref{z}), (\ref{w}), and  (\ref{solution}) give the
well known solution of the cubic equation (\ref{cubic}). The
question is how to choose a branch which tends to $s_1$ when $A_{i
j}\to 0$. The discriminant of this equation is
\begin{equation}
D\equiv \frac{q^2}{4} +\frac{p^3}{27}=d+\varepsilon
\end{equation}
 where
\begin{equation*}
d=\frac{1}{4}\Big
(-\frac{2}{27}\sigma_1^3+\frac{1}{3}\sigma_1\sigma_2-\sigma_3\Big )^2+
\frac{1}{27} \Big (\sigma_2-\frac{1}{3}\sigma_1^2\Big )^3
\end{equation*}
and
\begin{eqnarray*}
\varepsilon&=&\frac{1}{4}\delta^2+\frac{1}{2}\delta\Big
(-\frac{2}{27}\sigma_1^3+\frac{1}{3}\sigma_1\sigma_2-\sigma_3 \Big )\\
&-&
\frac{1}{27}\alpha^3+\frac{1}{9}\alpha^2(\sigma_2-\frac{1}{3}\sigma_1^2)-
\frac{1}{9}\alpha (\sigma_2-\frac{1}{3}\sigma_1^2)^2\ .
\end{eqnarray*}
Using the identity
\begin{equation*}
d=-\frac{1}{108}\Big [(s_1-s_2)(s_2-s_3)(s_3-s_1)\Big ]^2
\end{equation*}
the expression (\ref{solution}) can be transformed into
\begin{eqnarray*}
w&=& w_0-\frac{1}{2}\delta + \frac{{\rm i}}{6\sqrt{3}}(s_1-s_2)(s_2-s_3)(s_3-s_1)\\
&\times&\left [
\sqrt{1-\frac{108\varepsilon}{[(s_1-s_2)(s_2-s_3)(s_3-s_1)]^2}}-1\right ]
\end{eqnarray*}
where
\begin{eqnarray*}
w_0&=&\frac{1}{27}\sigma_1^3-\frac{1}{6}\sigma_1\sigma_2+\frac{1}{2}\sigma_3\\
&+&\frac{{\rm i}}{6\sqrt{3}}(s_1-s_2)(s_2-s_3)(s_3-s_1)\\
&=&\frac{1}{27}\Big (s_1+s_2{\rm e}^{-2\pi {\rm i}/3}+s_3{\rm e}^{2\pi {\rm i}/3}\Big )^3\ .
\end{eqnarray*}
Finally the root of the cubic equation (\ref{cubic}) which tends to
$s_1$ when $A_{i j}\to 0$ is
\begin{eqnarray*}
&&z=\frac{1}{3}s
\left [1-\frac{27\delta}{2s^3}+\frac{3\sqrt{3}{\rm i}}{2 s^3}(s_1-s_2)(s_2-s_3)(s_3-s_1)\right .\\
&&\times \left . \Big [
\sqrt{1-\frac{108\varepsilon}{[(s_1-s_2)(s_2-s_3)(s_3-s_1)]^2}}-1\Big ]\right ]^{1/3}
\end{eqnarray*}
where $s=s_1+s_2{\rm e}^{-2\pi {\rm i}/3}+s_3{\rm e}^{2\pi {\rm i}/3}$.

\end{document}